%% file: main.tex
\newif\ifPreprint
\Preprinttrue %

\ifPreprint
\documentclass[final]{agujournal2019preprint}
\else
\documentclass[draft]{agujournal2019}
\fi

\usepackage{comment}

\usepackage{lineno}

\ifPreprint\else
\linenumbers
\fi

\draftfalse
\journalname{Geophysical Research Letters}

\input{preamble.tex}

\begin{document}

\title{Real-time probabilistic tsunami forecasting in Cascadia from sparse offshore pressure observations}

\authors
{
Stefan Henneking\affil{1},
Fabian Kutschera\affil{3},
Sreeram Venkat\affil{1},\\
Alice-Agnes Gabriel\affil{3,4},
Omar Ghattas\affil{1,2}
}

\affiliation{1}{Oden Institute for Computational Engineering and Sciences, The University of Texas at Austin}
\affiliation{2}{Walker Department of Mechanical Engineering, The University of Texas at Austin}
\affiliation{3}{Scripps Institution of Oceanography, University of California San Diego}
\affiliation{4}{Department of Earth and Environmental Sciences, Ludwig‐Maximilians‐Universit\"at M\"unchen}

\correspondingauthor{Stefan Henneking}{stefan@oden.utexas.edu}

\ifPreprint\else
\begin{keypoints}
\item Real-time acoustic--gravity inversion accurately infers tsunamigenic seafloor motion from fully-coupled Cascadia dynamic rupture scenarios. %
\item A sparse network of 175 offshore seafloor pressure sensors supports accurate real-time tsunami forecasts in the Cascadia Subduction Zone.
\item Divergence of near-field seismo-acoustic wavefields after two minutes helps distinguish partial from margin-wide Cascadia ruptures.
\end{keypoints}
\fi

\ifPreprint
\fi

\begin{abstract} %
Near-field tsunami early warning in the Cascadia Subduction Zone is limited by sparse offshore observations. 
We investigate whether a hypothetical network of 175 ocean-bottom pressure sensors can support real-time Bayesian inference of the full spatiotemporal seafloor velocity field and probabilistic tsunami forecasting for a margin-wide and a partial fully-coupled Cascadia earthquake dynamic rupture–tsunami scenario. %
The simulated oceanic acoustic, Rayleigh, and tsunami wavefields %
are similar during the first two minutes after nucleation but diverge thereafter, enabling rapid earthquake scenario discrimination. 
Using an acoustic–gravity inversion %
with assimilation of pressure data, tsunami wave height forecasts are obtained in less than a second. 
We leverage a Bayesian inversion-based framework that splits the computations into an offline precomputation phase performed with large-scale computing facilities, and an online phase that computes forecasts %
and can be executed on a laptop. 
Forecast errors remain low at 22.1\% for the margin-wide and 19.6\% for the partial rupture. 
\end{abstract}

\section*{Plain Language Summary} %
	
Tsunamis generated by large offshore earthquakes can reach nearby coastlines within minutes, leaving little time for warning. 
In the Cascadia Subduction Zone, this challenge is made worse by a lack of offshore sensors to detect early signals.
This study explores whether a hypothetical network of 175 seafloor pressure sensors could improve near-field tsunami warnings in Cascadia. 
Using a fully-coupled earthquake--tsunami model, we simulate two realistic earthquake scenarios: a smaller, partial rupture and a larger, margin-wide rupture.
Synthetic pressure observations for two hypothesized sensor networks, one dense network with 600 sensors and one sparse network with 175 sensors, are used in an inversion framework that combines precomputed physics-based models with real-time data to rapidly estimate seafloor motion and predict tsunami wave heights.
The results show that accurate tsunami forecasts can be generated on a laptop in less than a second after receiving data. Despite the complexity of the wave signals, the system maintained relatively low forecast errors of about 20\%, with only a modest degradation for the sparser network.
These findings demonstrate that combining advanced modeling with expanded sensor networks could improve tsunami early warning systems, potentially saving lives by providing faster and more reliable tsunami forecasts for coastal communities.

\section{Introduction}
\label{sec:introduction}

\input{sections/introduction}

\section{Methods}
\label{sec:methods}
\input{sections/methods}

\section{Results}
\label{sec:results}
\input{sections/results}

\section{Discussion}
\label{sec:discussion}

\input{sections/discussion}

\section{Conclusions}
\label{sec:conclusions}

\input{sections/conclusions}

\section*{Open Research Section}
The open-source software SeisSol \cite<\url{https://seissol.org};>{Gabriel_2025_SeisSol}, available from GitHub (\url{https://github.com/SeisSol}, commit 1d03fba), was used for 3D dynamic rupture and fully coupled earthquake--tsunami modeling. The SeisSol input files required to run the simulations can be obtained from \url{https://doi.org/10.5281/zenodo.19141625} \cite{Henneking_2026_SupplementaryMaterialRealtime}. We incorporate the topography and bathymetry data of \citeA{Gebco_2020} using the following projection: \texttt{+proj=merc +lon\_0=231.43 +lat\_ts=46.8 +x\_0=-20082.185 +y\_0=-4069608 +datum=WGS84 +units=m}. All links were last accessed on May 24, 2026.

The open-source software MFEM \cite<\url{https://mfem.org};>{anderson2021mfem}, available from GitHub (\url{https://github.com/mfem}, commit 8921101), was used for acoustic--gravity model simulations.
The input files specifying sensor coordinates and tsunami forecasting coordinates, as well as the mesh file used for discretizing the seafloor velocity parameter field, can be obtained from \url{https://doi.org/10.5281/zenodo.19141625} \cite{Henneking_2026_SupplementaryMaterialRealtime}.

The open-source software FFTMatvec, available from GitHub (\url{https://github.com/s769/FFTMatvec}, commit 04282d7), was used for FFT-based matrix-vector products of the discretized parameter-to-observable maps in the acoustic--gravity model inversion framework.

\section*{Conflict of Interest declaration}
The authors declare there are no conflicts of interest for this manuscript.

\acknowledgments
Stefan Henneking, Fabian Kutschera, and Sreeram Venkat contributed equally to this work.

This research was supported by DARPA COMPASS HR0011-25-3-0242, DOD MURI grant FA9550-24-1-0327 and DOE ASCR grant DE-SC0023171.
FK and AAG acknowledge support from the National Science Foundation (NSF grant numbers EAR-2225286, EAR-2121568, OAC-2139536, OAC-2311208, RISE-2531036), the National Aeronautics and Space Administration (NASA 80NSSC20K0495), and the Statewide California Earthquake Center (SCEC grant 25341), as well as Horizon Europe (Geo-INQUIRE, grant number 101058518 and ChEESE-2P, grant number 101093038).

This research used resources from the National Energy Research Scientific Computing Center (NERSC; ALCC-ERCAP0030671, ScienceAtScale DDR-ERCAP0034808, NESAP DDR-ERCAP0038013). 
The authors acknowledge the Texas Advanced Computing Center (TACC) at The University of Texas at Austin for providing computational resources that have contributed to the research results reported within this paper. 
Additional computing resources were provided by the Institute of Geophysics of LMU Munich \cite{Oeser_2006_Cluster}.

\bibliography{main}

\clearpage

\ifPreprint\else
\makeatletter
\renewcommand{\ps@plain}{%
  \renewcommand{\@oddfoot}{\hfill\thepage} %
  \renewcommand{\@evenfoot}{\hfill\thepage} %
  \renewcommand{\@oddhead}{}%
  \renewcommand{\@evenhead}{}%
}
\makeatother

\pagestyle{plain}
\setcounter{page}{1}
\nolinenumbers
\fi

\begin{center}

\ifPreprint\else
\emph{Geophysical Research Letters}
\fi

\vspace{1ex}
Supporting Information for

\vspace{1ex}
\textbf{
Real-time probabilistic tsunami forecasting in Cascadia\\
from sparse offshore pressure observations
}

Stefan Henneking$^{1}$,
Fabian Kutschera$^{3}$,
Sreeram Venkat$^{1}$,\\
Alice-Agnes Gabriel$^{3,4}$,
Omar Ghattas$^{1,2}$

{
\small
$^{1}$Oden Institute for Computational Engineering and Sciences, The University of Texas at Austin\\
$^{2}$Walker Department of Mechanical Engineering, The University of Texas at Austin\\
$^{3}$Scripps Institution of Oceanography, University of California San Diego\\
$^{4}$Department of Earth and Environmental Sciences, Ludwig‐Maximilians‐Universität München\\
}

\end{center}

\vspace{1ex}
\noindent\textbf{Contents of this file}
\begin{itemize}
    \item Text S1 to S2
    \item Tables S1 to S3
    \item Figures S1 to S5
\end{itemize}

\ifPreprint\else
\vspace{1ex}
\noindent\textbf{Additional Supporting Information (Files uploaded separately)}
\begin{itemize}
    \item Captions for Movies S1 to S5
\end{itemize}
\fi

\vspace{1ex}
\noindent\textbf{Introduction}

\noindent
This supporting information provides additional text describing the 3D fully-coupled earthquake dynamic rupture and tsunami setup (Text~S1), as well as the 3D acoustic–gravity wave propagation simulation setup (Text~S2). We provide supporting figures for the analysis of the fully-coupled earthquake–tsunami simulations (Figures~S1--S5). In Table~S1, we summarize dynamic rupture parameters and key results for the two fully-coupled dynamic rupture scenarios expanded from \citeA{Glehman_2025_PartialRupturesGoverned}. Physical parameters used in the acoustic–gravity model are detailed in Table~S2, and the compute times for each phase of the Bayesian inversion framework are detailed in Table~S3.

\input{sections/supporting}

\end{document}

%% file: preamble.tex
\usepackage{colortbl}
\usepackage{physics}
\usepackage{mathtools}
\usepackage{booktabs}

\usepackage{cleveref}

\definecolor{RED}{RGB}{190, 30, 49}
\definecolor{BLUE}{RGB}{11, 78, 179}
\definecolor{GREEN}{RGB}{0, 171, 86}
\definecolor{OLIVE}{RGB}{93, 150, 72}
\definecolor{GREY}{RGB}{50, 50, 50}
\definecolor{BROWN}{RGB}{122, 90, 40}

\newcommand{\red}[1]{\textcolor{red}{#1}}

\newcommand{\mr}[1]{\mathrm{#1}}

\newcommand{\be}{\begin{equation}}
\newcommand{\ee}{\end{equation}}
\newcommand{\bes}{\begin{equation*}}
\newcommand{\ees}{\end{equation*}}

\newcommand{\p}{\partial}

\newcommand{\paramvec}{\vb{m}}
\newcommand{\datavec}{\vb{d}}

\newcommand{\qoivec}{\vb{q}}

\newcommand{\Kmat}{\mathbf{K}}
\newcommand{\Qmat}{\mathbf{Q}}
\newcommand{\ptomat}{\mathbf{F}}
\newcommand{\ptqmat}{\mathbf{F}_{\hskip -1pt q}}
\newcommand{\Gmat}{\mathbf{G}}

\newcommand{\numtime}{N_t}
\newcommand{\numparam}{N_m}
\newcommand{\numdata}{N_d}
\newcommand{\numqoi}{N_q}

\newcommand{\bfnoise}{\mathbf{\Gamma}_{\hskip -1pt \text{noise}}}
\newcommand{\bfpriorcov}{\mathbf{\Gamma}_{\hskip -1pt \text{prior}}}

\newcommand{\piprior}{\pi_{\text{prior}}}
\newcommand{\pipost}{\pi_{\text{post}}}
\newcommand{\pilike}{\pi_{\text{like}}}

\newcommand{\bfmmap}{\vb{m}_{\text{map}}}

\newcommand{\bfpostcovq}{\mathbf{\Gamma}_{\hskip -1pt \text{post}(q)}}

\newcommand{\bfqmap}{\vb{q}_{\text{map}}}

%% file: sections/introduction.tex
The most recent large megathrust earthquake on the Cascadia Subduction Zone (CSZ) occurred in 1700 \cite<e.g.,>{Atwater_1991_SuddenProbablyCoseismic,Satake_2003_FaultSlipSeismic}. Yet the rupture extent of past and future Cascadia earthquakes remains uncertain, with \mbox{(paleo-)}geological, geophysical, and modeling studies supporting both partial- and margin-wide earthquake rupture scenarios \cite{Walton_2021_IntegrativeGeologicalGeophysical, Melgar_2021_WasJanuary26th, Ramos_2021_AssessingMarginWideRupture, Small_2025_CombiningMultisiteTsunami, Glehman_2025_PartialRupturesGoverned, Goldfinger_2025_UnravellingDanceEarthquakes,DeSanto_2025_FullLockingShallow}.
Time-dependent recurrence modeling suggests a $\sim$30\% chance of a magnitude 8+ earthquake in southern Cascadia in the next 50 years, and $\sim$15\% probability of a magnitude $\sim$9 Cascadia-wide earthquake in the next 50 years \cite{Wirth_2025_EarthquakeProbabilitiesHazards}.

This uncertainty motivates numerical modeling to identify physically plausible scenarios consistent with past events such as the 1700 earthquake and tsunami. 
In Cascadia, coastal paleo-subsidence estimates provide key constraints on modeled slip heterogeneity and tsunami hazard \cite<e.g.,>{Atwater_1995_SummaryCoastalGeologic, Wang_2013_HeterogeneousRuptureGreat, Kemp_2018_RevisingEstimatesSpatially, Salaree_2021_RelativeTsunamiHazard, Dunham_2025_Impact3DStructure}.
Dynamic rupture simulations provide a physics-based framework to test which rupture scenarios are mechanically viable under realistic initial conditions \cite<e.g.,>{Day_1982_ThreedimensionalSimulationSpontaneous, Olson_1997_3D_Landers, Ramos_2022_IntroDR}.

Fully-coupled earthquake–tsunami models combine earthquake scenarios with acoustic wave propagation and tsunami gravity wave generation.
This method was introduced by \citeA{Maeda_2013_FDMSimulationSeismic} and applied by \citeA{Maeda_2013_SeismicTsunamiWave} to the 2011 Tohoku--Oki earthquake and tsunami using a kinematic source model.
For Cascadia, fully-coupled dynamic rupture and tsunami simulations have previously been carried out using two-dimensional models \cite{Lotto_2018_AppliedFullyCoupled, Wilson_Ma_2021_FullyCoupledDR}.
However, recent 3D dynamic rupture simulations constrained by Cascadia paleo-subsidence and kinematic geodetic coupling models 
\cite{Ramos_2021_AssessingMarginWideRupture, Glehman_2025_PartialRupturesGoverned, Biemiller_2025_StructuralControlsSplay} have not yet been extended to model the ensuing tsunami.

Effective tsunami early warning and detection require multiple techniques and data sources. One established method based on direct hydrodynamic measurements involves Deep-ocean Assessment and Reporting of Tsunamis stations \cite<DART;>{Gonzalez_1998_DeepoceanAssessmentReporting}, which are part of some operational real-time tsunami forecasting systems \cite<e.g.,>{Wei_2008_RealtimeExperimentalForecast, Mungov_2013_DARTTsunameterRetrospective, Tang_2017_RealTimeAssessment16}.
Alternative systems are based on onshore seismic and geodetic observations \cite<e.g.,>{Tsuboi_1995_RapidDeterminationMw, Kanamori_2008_SourceInversionOfWphasea, Gusman_2014_PhaseInversionTsunami, Crowell_2016_DemonstrationCascadiaGFAST, Williamson_2020_NearFieldTsunamiForecasting}, and combined seismo-geodetic approaches \cite<e.g.,>{Melgar_2013_FieldTsunamiModels, Golriz_2023_RealTimeSeismogeodeticEarthquake}. 
Such rapid earthquake source parameter estimation is currently the basis for tsunami early warnings issued by U.S. Tsunami Warning Centers \cite{Hirshorn_2020_EarthquakeSourceParameters}. 

However, the sparsity of available sensors, particularly offshore, still limits the ability to constrain tsunami generation close to the source. 
Following the 2011 magnitude 9 Tohoku–Oki earthquake and tsunami, Japan deployed offshore sensor networks, including DONET and S-net.
The latter consists of an array of 150 ocean-bottom stations, each equipped with pressure gauges and seismometers, interconnected by about 5800~km of fiber-optic cable to improve earthquake and tsunami monitoring \cite<e.g.,>{Kanazawa_2013_JapanTrenchEarthquake, Tanioka_2018_NearfieldTsunamiInundation}. 
These networks may enable rapid tsunami detection and forecasting from near-fault ocean-bottom pressure data \cite{Aoi_2020_MOWLASNIEDObservation, Mizutani_2020_EarlyTsunamiDetection, Mulia_2021_SyntheticAnalysisEfficacy}. 
Similar offshore sensor networks for earthquake and tsunami early warning have been proposed for Cascadia~\cite{Saunders_2018_AugmentingOnshoreGNSS, schmidt2019monitoring}.
At the same time, distributed acoustic sensing (DAS) %
is expanding offshore observational coverage and may improve our ability to resolve 
tsunami characteristics \cite{Xiao_2024_DetectionEarthquakeInfragravity, Tonegawa_2024_HighFrequencyTsunamisExcited, Becerril_2026_TsunamiEarlyWarningDistributed}.

In this study, we demonstrate that a Cascadia offshore network comparable in scale to S-net could enable accurate real-time probabilistic tsunami forecasting. We utilize a hypothetical network of 175 ocean-bottom pressure sensors to drive a physics-based inversion framework. 
Prior work using near-field ocean-bottom pressure data for tsunami source characterization has typically employed low-dimensional parameterizations of earthquake source parameters, such as moment magnitude, hypocenter location, or simplified static slip distributions, matched either against precomputed scenario databases or against long-wave shallow-water forward models \cite{Tsushima2009,Tanioka_2018_NearfieldTsunamiInundation, Mulia_2021_SyntheticAnalysisEfficacy, Kubota_2021_ExtractingNearFieldSeismograms}.
Other work bypassed source estimation by assimilating pressure data into a shallow-water tsunami propagation model, updating the tsunami wavefield without reconstructing its source \cite{Maeda2015,Gusman2016}. 
Unlike these approaches, our framework inverts directly for the full spatiotemporal seafloor velocity field, without assuming a causative source fault geometry or dislocation model, using a 3D coupled near-field acoustic--gravity wave propagation model that captures the complex seafloor motion immediately following megathrust rupture nucleation.

This study extends the methodology and results previously presented in~\citeA{henneking2025bell}. Whereas that work demonstrated the computational scalability of the inverse solver for an idealized sensor configuration, this paper focuses on increased physical realism and application relevance. The key differences are as follows:
\begin{enumerate}
    \item \textbf{More realistic source physics:} we transition from the solid Earth-only dynamic rupture scenarios \cite{Glehman_2025_PartialRupturesGoverned} used in~\citeA{henneking2025bell} to fully-coupled dynamic rupture--tsunami models that include the interaction between the solid Earth and the ocean water layer. These fully-coupled models provide a more physically consistent representation of the evolving seafloor motion during rupture and the energy transfer to the water column. 
    \item \textbf{Variable rupture scenarios:} whereas the previous work considered only one margin-wide rupture scenario,  
    we additionally introduce a partial rupture scenario. 
    This additional scenario allows us to assess whether our system can distinguish between earthquakes of varying spatial extents. 
    \item \textbf{Realistic sensor network size:} the inversion in~\citeA{henneking2025bell} used an array of 600 sensors spaced near-uniformly across the CSZ. Here, we instead perform the inversion using a much sparser network of 175 sensors, chosen to reflect a plausible deployment size comparable to S-net in Japan.
\end{enumerate}

%% file: sections/methods.tex
\subsection{Cascadia fully-coupled dynamic rupture and tsunami simulations}

\label{subsec:dynamic-rupture-simulations}

\begin{figure}[htbp]
    \centering
    \includegraphics[width=\textwidth,trim={0 0.15in 0 0}]
    {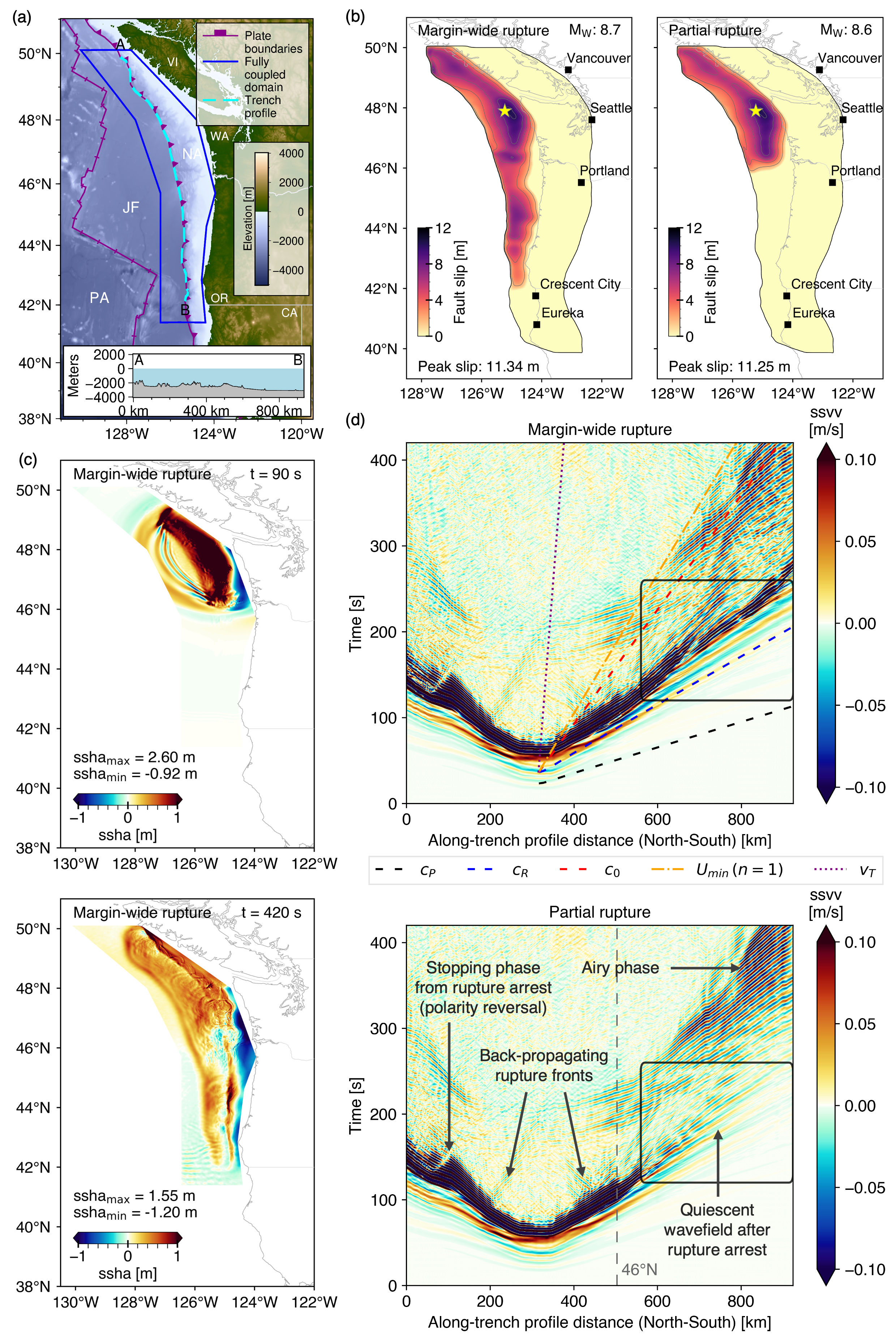}
    \caption{\footnotesize (a) Overview of the Pacific Northwest coast and major plate boundaries \cite{Bird_2003_UpdatedDigitalModel}. The blue polygon outlines the domain for the 3D fully-coupled earthquake dynamic rupture and tsunami simulations performed with SeisSol \cite{Gabriel_2025_SeisSol}. The dashed cyan line marks a densely spaced virtual receiver profile located at the sea surface, following the Juan de Fuca (JF) and North American (NA) plate boundary from 50°N to 42°N. 
    (b) Earthquake slip distributions for the margin-wide and partial rupture fully-coupled simulations, 
    which reproduce the solid-Earth dynamic rupture scenarios of \citeA{Glehman_2025_PartialRupturesGoverned} with an added ocean water layer. 
    (c) Sea surface height anomaly (ssha) for the margin-wide rupture scenario at 90~s, during tsunami generation, and at 420~s, at the end of the fully-coupled simulation. Corresponding snapshots for the partial rupture scenario are shown in Figure~S2.
    (d) Annotated space-time evolution of sea surface vertical velocities (ssvv) along the trench profile for the margin-wide and partial rupture scenarios. The dashed lines mark the respective velocities of seismic P-waves ($c_P$; black), solid Earth Rayleigh waves ($c_R$; blue), sound waves in water ($c_0$; red), as well as the minimum group velocity of mode $n=1$ oceanic Rayleigh waves ($U_{\min}$; orange) following \citeA{Abrahams_2023_ComparisonMethodsCoupled}, and the average tsunami wave speed ($v_T$; purple). 
    The gray vertical dashed line marks $46^\circ$N, near the arrest location of the southward-propagating partial rupture.
    }
    \label{fig:figure01}
\end{figure}

We extend two Cascadia 3D dynamic rupture scenarios from \citeA{Glehman_2025_PartialRupturesGoverned} to simulate fully-coupled earthquake–tsunami generation. We consider a partial rupture ($M_W$~8.6) and a margin-wide rupture ($M_W$~8.7) scenario, corresponding to their Models 2 and 16 (Supporting Information, Table~S1). Each model combines the Slab2 Cascadia megathrust interface \cite{Hayes_2018_Slab2ComprehensiveSubduction} with realistic bathymetry and topography \cite{Gebco_2020}. Both use a linear slip-weakening friction law \cite<e.g.,>{Ida_1972_CohesiveForce_LSW,Andrews_1976_RuptureVelocity_LSW}, with a static friction coefficient of 0.6, a dynamic friction coefficient of 0.1, and a critical slip-weakening distance of 1.0~m, following \citeA{Ramos_2021_AssessingMarginWideRupture}. Both scenarios also assume the same hypocenter in northern Cascadia at 16~km depth, a pore fluid pressure ratio of 0.97 near lithostatic conditions \cite{Madden_2022_StatePoreFluid}, and a low rigidity structure \cite{Sallares_2019_UpperplateRigidityDetermines}.  
The two models differ in their initial stress conditions. The partial rupture scenario has lower prestress in central Cascadia, creating a stress barrier that arrests rupture.  
The addition of acoustic and gravity wave propagation within the added ocean layer (Figure~\ref{fig:figure01}a) does not alter rupture dynamics and slip distribution in the margin-wide and partial rupture scenarios (Figure~\ref{fig:figure01}b) compared to \citeA{Glehman_2025_PartialRupturesGoverned}'s dynamic-rupture-only results. 

We use the open-source software SeisSol \cite{Gabriel_2025_SeisSol} for the fully-coupled earthquake dynamic rupture and tsunami simulations (Supporting Information, Text~S1). 
We add a 3D acoustic layer representing the ocean atop the solid Earth, with shear modulus set to zero, and use a uniform density. %
Together with the underlying elastic structure, this configuration governs the modeled seismic, acoustic, and oceanic Rayleigh wave propagation. To model gravity tsunami waves, we apply a gravity-restoring boundary condition at the sea surface, adopting an Eulerian description \cite{Lotto_2015_HighorderFiniteDifference,Krenz_2021_FullyCoupled,Abrahams_2023_ComparisonMethodsCoupled}. We define the sea surface height anomaly (ssha, Figure~\ref{fig:figure01}c) as the deviation from the ocean surface at rest. 
For the Bayesian inversion framework described in the next section, we sample the modeled seafloor normal velocities at 1~Hz (Supporting Information, Text~S1) and use these time histories as prescribed boundary inputs to the acoustic--gravity model, which then generates the synthetic ocean-bottom pressure observations used for assimilation.

\subsection{Bayesian inversion framework}
\label{subsec:bayesian-inversion-framework}

Our approach combines fully-coupled earthquake–tsunami simulations of the Cascadia rupture scenarios (\Cref{subsec:dynamic-rupture-simulations}) with an acoustic–gravity model used to infer tsunamigenic seafloor motion and forecast tsunamis from seafloor pressure observations. 
The fully-coupled simulations are used to generate physically-consistent synthetic rupture scenarios. The Bayesian inversion and tsunami forecasting are performed with the acoustic–gravity model.
We note that the fully-coupled simulations serve only as a source of physically-consistent synthetic seafloor motion, but the mapping from this seafloor motion to ocean-bottom pressure is linear under the acoustic–gravity model, independent of how the seafloor motion itself is generated.
This choice enables us to resolve near-field coupled acoustic and gravity wave dynamics driven by the full spatiotemporal seafloor motion while retaining a computationally efficient forward and adjoint framework for real-time inference.
Our Bayesian inversion assimilates the complete time history of synthetic seafloor pressure observations over the 420-second simulation window, capturing both the early high-frequency acoustic arrivals and the evolving longer-period pressure changes associated with the developing seafloor deformation and gravity wave generation. 

The coupled acoustic--gravity wave equations model pressure perturbations and tsunami propagation in a compressible ocean, thereby capturing interactions between surface gravity waves and acoustic waves. The partial differential equations (PDEs) are derived by linearizing the mass and momentum conservation laws about a hydrostatic background state. Surface gravity waves enter through an adjusted kinematic free-surface boundary condition~\cite{Lotto_2015_HighorderFiniteDifference}. 
The governing equations take the form of a coupled first-order system involving the fluid velocity $\vec{u}(\vec{x},t)$, acoustic pressure $p(\vec{x},t)$, and surface wave elevation $\eta(\vec{x},t)$,
\begin{equation}
        \left\{
        \begin{aligned}
                \rho\, \p_t \vec{u} + \nabla p &= 0, & \Omega \times (0,T), \\
                K^{-1} \p_t p + \nabla \cdot \vec{u} &= 0, & \Omega \times (0,T), \\
                p &= \rho g \eta, & \p \Omega_{\text s} \times (0,T), \\
                \p_t \eta &= \vec{u} \cdot \vec{n}, & \p \Omega_{\text s} \times (0,T), \\
                \vec{u} \cdot \vec{n} &= -m, & \p \Omega_{\text b} \times (0,T), \\
                \vec{u} \cdot \vec{n} &= Z^{-1} p, & \p \Omega_{\text a} \times (0,T), \\
        \end{aligned}
        \right.\label{eq:fwd-pde}
\end{equation}
subject to zero initial state conditions. In this formulation, $\rho$ and $K$ denote the water's density and bulk modulus, respectively, defining the sound speed $c = \sqrt{K/\rho}$ and acoustic impedance $Z = \rho c$. The variable $g$ represents gravitational acceleration. The parameters used in our model are specified in Supporting Information, Table~S2. The free surface $\p \Omega_{\text s}$, the ocean floor $\p \Omega_{\text b}$, and the artificial lateral absorbing boundaries $\p \Omega_{\text a}$ comprise the boundary of the compressible ocean domain $\Omega$. The vector $\vec{n}$ denotes the outward-pointing unit normal, and the temporal domain is $(0,T)$.
The forward model inputs are given by $m(\vec{x},t)$, a spatiotemporal field that prescribes the (earthquake-induced) inward-pointing normal velocity of the seafloor.

The observables %
are the model predictions of ocean-bottom pressures $\mr{d}^\text{obs}$ obtained from seafloor sensors. 
The \emph{quantities of interest} (QoI), denoted by $\mr{q}$, represent model predictions of tsunami wave heights at target locations and times.
Given seafloor pressure observations $\mr{d}^\text{obs}$, the goal of the inversion framework is to (1)~infer the uncertain (discretized) seafloor velocity parameters $\mr{m}$, and (2)~use the inferred parameters $\mr{m}$ as inputs to the (discretized) acoustic--gravity equations to forecast tsunami wave heights $\mr{q}$.

The challenge with solving this inverse problem, in real time, is the computational cost of solving the acoustic--gravity model at high-enough resolution to accurately resolve the propagating ocean acoustic waves.
While earthquake-induced seafloor motion contains a range of frequencies, the dominant acoustic wave frequencies are typically of the order of $\sim$0.1~Hz in the deep ocean \cite{leveque2018cascadia, Levin_2015_PhysicsTsunamis}.
The coupled acoustic–gravity model resolves a spectrum of frequencies from the long-period gravity wave regime up to 1~Hz (i.e., 1500~m wavelengths) for the propagating acoustic waves. Resolving this across the 1000~km long CSZ in the 3D compressible ocean domain requires fine spatial grids. 
For Cascadia, the discretized spatiotemporal seafloor velocities $\mr{m}$ consist of over one billion parameters.
Moreover, the inference of this high-dimensional parameter vector from sparse and noisy pressure data is inherently ill-posed, as many different sets of parameters may be consistent with the limited observational data to within the noise tolerance.

To address this challenge, we adopt a Bayesian approach. Rather than seeking a single deterministic ``best-fit'' solution, Bayesian inference characterizes the probability that any given parameter field is consistent with both the observed data and any prior knowledge. The uncertain parameter field is treated as a random field, and the solution to the inverse problem is the posterior probability distribution~\cite{KaipioSomersalo05, stuart2010inverse, ghattas2021learning}.
This probabilistic framework rigorously quantifies uncertainties in both the inverse solution and the resulting tsunami forecasts.

For the Cascadia inversion, the Bayesian solution characterizes the \textit{posterior} probability density of the discretized seafloor velocities $\mr{m}$, conditioned on the pressure observations $\mr{d}^\text{obs}$. The subsequent predictive task propagates this inferred posterior through the forward model to obtain the \textit{posterior predictive} of the tsunami wave heights $\mr{q}$. In finite dimensions, Bayes' rule states
\bes
	\pipost(\mr{m} | \mr{d}^\text{obs}) \propto \pilike(\mr{d}^\text{obs} | \mr{m}) \, \piprior(\mr{m}) ,
\ees
where $\pipost(\mr{m} | \mr{d}^\text{obs})$ denotes the posterior probability density of the parameters $\mr{m}$ given observations $\mr{d}^\text{obs}$; $\pilike(\mr{d}^\text{obs} | \mr{m})$ is the likelihood of observations, which measures data misfit; and $\piprior(\mr{m})$ is the prior, which encapsulates any initial knowledge about the parameters.

The acoustic–gravity forward map and the forecast quantity-of-interest map are linear. Combined with a Gaussian prior and an additive Gaussian noise model, where observed pressure signals are assumed as the sum of the true signals and Gaussian-distributed noise, the posteriors of the seafloor velocity parameters and the tsunami forecasts are also Gaussians, implying that the parameter inference reduces to a linear inverse problem.
Nonetheless, the high dimensionality of the parameter space, the computational complexity of the discretized PDE model, and the wave-propagation character of the governing equations preclude conventional methods solving this linear inverse problem in real time~\cite{henneking2025bell}.

Instead, we employ a recently developed framework that exploits the linear time-invariant structure of the acoustic--gravity wave equations~\cite{henneking2026goal}.
In this framework, the inverse solution is decomposed into several precomputation (offline) phases that are executed just once, before any earthquake occurs, and a real-time (online) phase of parameter inference and tsunami wave height forecasts that is executed when an earthquake occurs and data are acquired. 
The offline phases are computationally demanding and require performance-optimized implementations on large-scale computing systems~\cite{tu2026tensor}. 
The framework's most expensive step is precomputing one adjoint PDE solution of the acoustic--gravity model for each sensor location and each forecast location. 
The acoustic--gravity PDEs are discretized using high-order finite element methods (Supporting Information, Text~S2) and implemented via the MFEM library~\cite{anderson2021mfem}. 
For Cascadia, each wave propagation solution takes approximately 52~minutes on 512 GPUs (Supporting Information, Table~S3).

Once these precomputations are completed, the remaining computations, both offline and online, do not require solving PDEs but instead exploit the time invariance property of the underlying physics model to perform fast and efficient FFT-based mappings from the parameter inputs to the observables~\cite{venkat2025fft}.
In the online phase, given real-time data from seafloor sensors, the seafloor motion and tsunami wave heights are computed within fractions of a second (0.2~s for seafloor motion and 1~ms for tsunami wave heights at 21 forecasting locations, Supporting Information, Table~S3).
For a detailed discussion of this real-time Bayesian inference framework, we refer to \citeA{henneking2026goal} and references therein.

%% file: sections/results.tex
\subsection{Fully-coupled margin-wide and partial dynamic rupture models}
\label{subsec:results-dynamic-rupture}

The margin-wide and partial rupture scenarios nucleate as crack-like ruptures and then evolve into bilateral, sub-Rayleigh, pulse-like ruptures \cite{Glehman_2025_PartialRupturesGoverned}. 
They exhibit similar rupture dynamics until the respective southward rupture front reaches the central CSZ near the Oregon-Washington border (Supporting Information, Figure~S1, Movie~S1). 
The margin-wide rupture propagates further southward and arrests at $\sim$42°N near the Oregon-California border, whereas the partial rupture scenario arrests at $\sim$46°N (Figure~\ref{fig:figure01}b), remaining constrained to offshore Vancouver Island and Washington.
The northward rupture fronts in both scenarios propagate to the northern end of the megathrust geometry, where they terminate abruptly. 
The rapid deceleration and cessation of slip radiate stopping phases \cite{Savage_1964_PropertiesTensileFractures, Savage_1965_StoppingPhaseSeismograms, Madariaga_1976_DynamicsExpandingCircular, Schliwa_2023_EquivalentNearFieldCorner, Kearse_2026_StoppingPhaseReveals}. 
Both rupture scenarios exhibit low-slip-rate back‐propagating rupture fronts visible between $\sim$90 to 120~s  \cite<e.g.,>{Gabriel_2012_TransitionDynamicRupture, Ding_2024_BackPropagatingRuptureNature, Sun_2026_BackPropagatingEarthquakesSimple}, due to prestress heterogeneity in central Cascadia and the geometric barrier imposed by the finite slab geometry in northern Cascadia (Supporting Information, Figure~S1, Movie S1).
Despite their rupture complexity, the two scenarios have comparable peak slip values of 11.34~m and 11.25~m, as well as similar average slip, peak slip rate, and rupture speed (Supporting Information, Table~S1). 

The early acoustic and tsunami response mirrors the initially similar dynamic rupture evolution (Movie~S2). For example, both fully-coupled simulations produce comparable ssha extrema of $\sim$2.5~m and $\sim$-1.0~m at 60~s simulation time (Supporting Information, Figure~S2). 
The wavefields diverge once the partial rupture arrests near 46°N, while the margin-wide rupture continues southward and keeps generating seismo-acoustic and gravity waves across central Cascadia (Figure~\ref{fig:figure01}c; Supporting Information, Figure~S2). 
Over the simulation duration, peak tsunami amplitudes reach $\sim$4~m and $\sim$2.9~m for the margin-wide and partial rupture scenario, respectively.

To compare the space-time evolution of sea surface vertical velocities (ssvv; Figure~\ref{fig:figure01}d), we place a line of synthetic receivers at the sea surface, following the trench from 50°N to 42°N over a distance of 923~km (Figure~\ref{fig:figure01}a; Supporting Information, Figure~S3). 
The modeled ocean wavefield contains multiple seismo-acoustic phases in addition to the long-period, dispersive tsunami gravity wave signal. 
The fastest, low-amplitude sea surface signals arise from seismic P-waves ($c_P$) and S-waves, which dynamically perturb the seafloor and overlying water column and thereby produce sea-surface motion preceding the developing tsunami signal \cite<e.g.,>{Saito_2019_SynthesizingSeaSurface, Wirp_2021_3DLinkedSubduction}.

Oceanic Rayleigh waves propagate slower than seismic body and solid-Earth Rayleigh waves \cite<e.g.,>{Biot_1952_InteractionRayleighStoneley,Nakanishi_1992_RayleighWavesGuided,Kozdon_2014_ConstrainingShallowSlip,Abrahams_2023_ComparisonMethodsCoupled}.
These interface waves are guided by the coupled solid Earth-ocean system and are dispersive \cite<e.g.,>{Noguchi_2016_OceaninfluencedRayleighWaves}, with velocities spanning the solid Earth Rayleigh-wave speed ($c_R$), the acoustic wave speed in water ($c_0$), and the mode-1 minimum group velocity ($U_{\min}$; Supporting Information, Figure~S4). 
In Figure~\ref{fig:figure01}d, $c_R$ is assumed to be 0.92 times the shear-wave speed, calculated with TauP and ObsPy \cite{Crotwell_1999_TauPToolkitFlexible,Beyreuther_2010_ObsPy} using the same 1D solid-Earth velocity structure assumed in the dynamic rupture simulations \cite{Sallares_2019_UpperplateRigidityDetermines, Glehman_2025_PartialRupturesGoverned}.

The late, high-amplitude dispersive sea surface wave packet visible in southern Cascadia for both rupture scenarios is the Airy phase \cite{Press_1950_AiryPhaseShallowfocus}, associated with the mode-1 group-velocity minimum and controlled by water-column depth \cite<e.g.,>{Pekeris_1948_TheoryPropagationExplosive,Abrahams_2023_ComparisonMethodsCoupled}. 
The slowest signals are tsunami gravity waves, whose sea surface vertical velocity amplitudes remain small along the north–south profile. Along this profile, the mean water depth is 2554~m. 
Linear shallow-water theory \cite<e.g.,>{Talley_2011_ChapterGravityWaves, Levin_2015_PhysicsTsunamis, Kundu_2016_ChapterGravityWaves} predicts an average tsunami speed ($v_T$) of $158$~m~\!s$^{-1}$. 

After identifying these propagation phases, we use the ssvv space--time profiles to isolate dynamic rupture source effects at the sea surface. Once the southward-propagating partial rupture arrests near 46°N, the high-amplitude dynamic rupture-driven vertical sea surface velocity signals decay by about 80\%, providing a proxy for the along-strike extent of active rupture. 
After this arrest, the oceanic Rayleigh wavefield is no longer overprinted by continuing rupture-generated signals, unlike in the margin-wide scenario (Figure~\ref{fig:figure01}d; Supporting Information, Figure~S5, Movie~S3). 
Coseismic back-propagating rupture fronts appear as faint arrivals with slopes opposite to the dominant southward rupture direction in the ssvv space--time profiles (Figure~\ref{fig:figure01}d). The low-amplitude ``gap'' near the northern end of the profiles is due to destructive interference between the reversed-polarity stopping phase and the preceding wavefield.  

\subsection{Inference of seafloor motion and forecast of tsunami wave heights}

The complex ground motions generated by the fully-coupled earthquake dynamic rupture and tsunami simulations of \Cref{subsec:results-dynamic-rupture} are used to evaluate our real-time inversion methodology for Cascadia.
Treating these ground motions as ``ground truths'' for the seafloor velocity parameter field, we use the forward acoustic--gravity solver to generate synthetic pressure observations at hypothetical sensor locations, which are sampled at 1~Hz (Supporting Information, Text~S2), for each rupture scenario.
We add 1\% relative Gaussian noise to the pressure signals and compare two hypothetical sensor arrays consisting of $\numdata=600$ versus $\numdata=175$ pressure sensors. 
Given these synthetic observational data, we use the Bayesian inversion framework described in~\Cref{subsec:bayesian-inversion-framework} to infer the earthquake-induced seafloor motion in the CSZ and forecast tsunami wave heights at $\numqoi=21$ designated locations.

\begin{figure}[htbp]
    \centering
    \includegraphics[width=\textwidth]
    {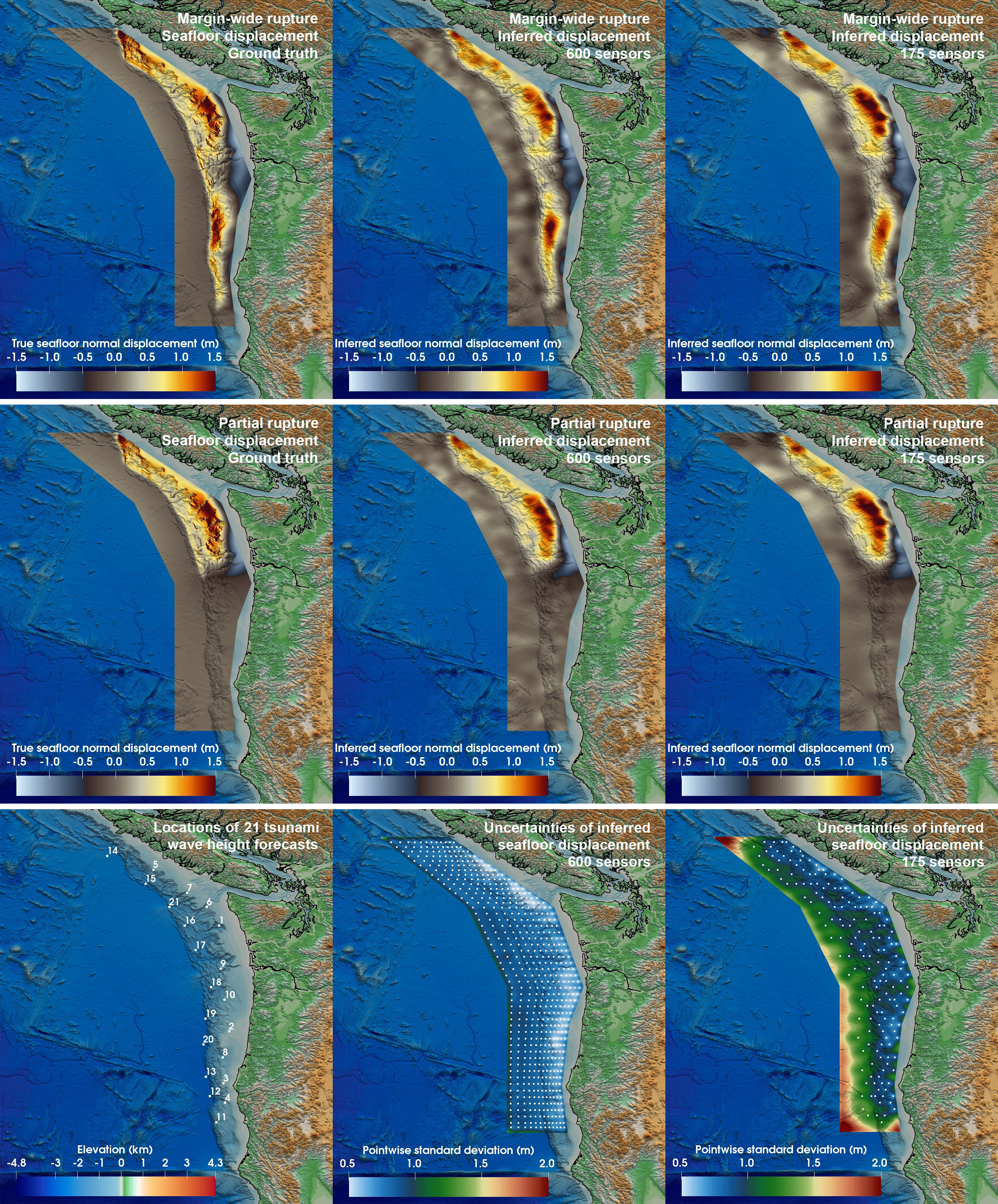}
    \caption{Snapshots of the true (left column) and inferred (middle and right columns) seafloor normal displacements at the final time ($t$~=~420~seconds) for a margin-wide rupture (top row) and partial rupture (middle row) scenario, and their associated uncertainties given as pointwise standard deviations (bottom row). Results in the middle and right columns respectively correspond to inversion using synthetic data from 600 and 175 hypothesized ocean bottom pressure sensors, with the sensor locations marked in the corresponding uncertainty plots of the bottom row. The bottom left plot depicts ocean bathymetry with the 21 tsunami wave height forecasting locations.}
    \label{fig:inversion-results}
\end{figure}

\begin{figure}[htbp]
    \centering
    \includegraphics[width=\textwidth]
    {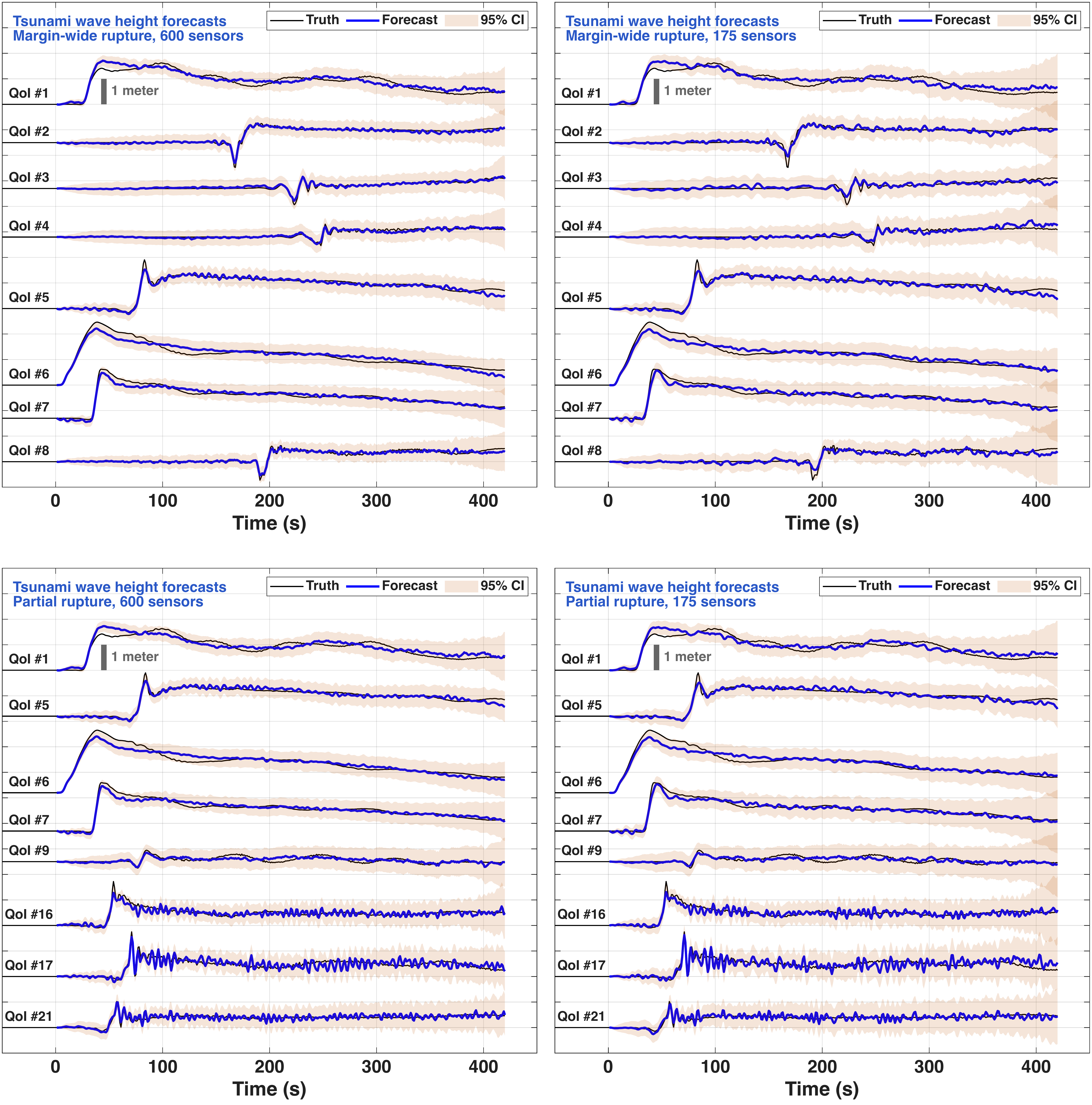}
    \caption{Tsunami wave height forecasts and their uncertainties (depicted as 95\% credible intervals) at a subset of the 21 target locations (labeled as QoI \#1--\#21, with locations as shown in the bottom left plot of Figure~\ref{fig:inversion-results}), obtained with 600 sensors (left column) and 175 sensors (right column), for the margin-wide rupture (top row) and partial rupture scenarios (bottom row).
    The relative errors of the tsunami forecasts are 18.6\% (600 sensors) and 22.1\% (175 sensors) for the margin-wide rupture, showing only a modest degradation of the forecast quality with the sparser sensor network. Similar tsunami forecast errors of 18.1\% (600 sensors) and 19.6\% (175 sensors) are obtained for the partial rupture scenario.}
    \label{fig:forecast-results}
\end{figure}

The accuracy of (the mean of) the inferred parameters and tsunami wave height forecasts can be estimated by computing their relative errors (measured as l2 vector norms) against the ground truths.
For the margin-wide rupture scenario, the relative errors of the inferred (spatiotemporal) seafloor normal displacement field are 29.6\% (600 sensors) and 36.1\% (175 sensors); and the relative errors of the tsunami wave height forecasts are 18.6\% (600 sensors) and 22.1\% (175 sensors).
For the partial rupture scenario, the relative errors of the inferred seafloor displacements are 27.7\% (600 sensors) and 31.7\% (175 sensors); and the relative errors of the tsunami forecasts are 18.1\% (600 sensors) and 19.6\% (175 sensors).

The results of the parameter inference and tsunami wave height forecasts under uncertainty 
are shown in Figures~\ref{fig:inversion-results} and \ref{fig:forecast-results} for both the margin-wide and the partial rupture scenario.
The ``ground truths'' generated by the fully-coupled dynamic rupture–tsunami simulations and (the means of) the inverse solutions in Figure~\ref{fig:inversion-results} are depicted in terms of the final seafloor normal displacement, $\int_0^T m(\vec x, t) dt$, at $T=420$~s.
Animations showing the temporal evolution of the true and inferred displacement fields are provided in Movies S4 \& S5.
Similarly, the associated uncertainties of the Bayesian inverse solutions are shown as pointwise standard deviations of the final seafloor normal displacements in Figure~\ref{fig:inversion-results}.
The tsunami wave height forecasts in Figure~\ref{fig:forecast-results} are in each case (top row: margin-wide rupture; bottom row: partial rupture) shown for a subset of the 21 target forecasting locations depicted in the bottom left plot of Figure~\ref{fig:inversion-results}. 
The forecasts are compared against the synthetic truths generated by the acoustic--gravity model using the ``ground truth'' seafloor velocity fields from the dynamic rupture models as an input.
The time-dependent uncertainties associated with each forecast are illustrated as 95\% credible intervals (CIs) around the mean of each forecast.

The inverse solutions show that a sparse network of 175 offshore pressure sensors supports accurate real-time tsunami forecasts in the CSZ.
The forecasts with 175 sensors are nearly as accurate as the ones obtained with 600 sensors, in both the margin-wide rupture and the partial rupture scenario.
The inferred seafloor displacements have larger errors than the tsunami forecasts, and Figure~\ref{fig:inversion-results} shows that these errors are primarily due to the inability of inferring the fine spatial scales of the seafloor uplift. 
Thus, while sparse offshore pressure observations in the near field are not sufficient to recover all components of the displacement field accurately, the \emph{tsunamigenic} components of the seafloor motion are recovered with good accuracy.
For both the seafloor displacements and the tsunami forecasts, the inverse solution with 175 sensors clearly distinguishes the margin-wide rupture from the partial rupture scenario.

%% file: sections/discussion.tex
\subsection{Implications for near-field tsunami forecasting}

The fully-coupled earthquake dynamic rupture and tsunami models illustrate that the near-field ocean wavefield contains not only the emerging tsunami signal but also faster seismo-acoustic phases, including oceanic Rayleigh waves and acoustic waves that may accompany submarine earthquakes \cite<e.g.,>{Nakamura_2012_FDMSimulationSeismicWave,Kozdon_2014_ConstrainingShallowSlip,Abdolali_2015_HydroacousticTsunamiWaves}. 
These phases are also present in the ocean-bottom pressure observations assimilated by the inversion framework. 
In the margin-wide and partial rupture scenario considered here, these phases are excited similarly during the first two minutes of rupture propagation, limiting discrimination potential between partial and margin-wide rupture during the earliest part of the records. 

After rupture arrest in the partial scenario, however, the high-amplitude dynamic rupture-driven signals decay rapidly, whereas they continue in the margin-wide scenario. This divergence suggests that early seismo-acoustic wavefields may encode rupture extent before the long-period, slower tsunami signal becomes diagnostic, with implications for operational near-field tsunami warning \cite<e.g.,>{Wang_2012_RealtimeForecastingApril, Melgar_2016_LocalTsunamiWarnings}. 
This interpretation is consistent with recent observational and modeling studies showing that coherent stopping phases are most efficiently generated when ruptures arrest abruptly near rupture endpoints, whereas gradual deceleration or self-arrest may produce weaker or absent stopping phases \cite<e.g.,>{Kame_2003_DynamicBranchingArresting, Xu_2023_2016MenyuanEarthquake, Schliwa_2023_EquivalentNearFieldCorner, Mosconi_2026_DiscriminatingDynamicRupture, Kearse_2026_StoppingPhaseReveals}. 
In our simulations, this distinction is reflected by stopping phases from the abrupt northern rupture termination, but by a rapid decay of high-amplitude rupture-driven sea surface velocity signals after the southward partial rupture arrests near 46°N.

Future work could extend our analysis to other rupture scenarios, including events nucleating in southern Cascadia or exhibiting different along-strike prestress or frictional segmentation \cite{Ramos_2021_AssessingMarginWideRupture,Wirth_2025_EarthquakeProbabilitiesHazards,Small_2025_CombiningMultisiteTsunami}. 
Among these possible extensions, uncertainty in the prescribed fault geometry and subsurface structure is likely particularly important because both coseismic seafloor deformation and tsunami generation depend on the assumed megathrust interface and velocity model \cite{Witter_2013_SimulatedTsunamiInundation, Lotto_2018_AppliedFullyCoupled, Kwong_2025_PerformanceSlabGeometry, Dunham_2025_Impact3DStructure, Wirth_2025_ThreedimensionalSeismicVelocity}. The dynamic rupture models considered here use the Slab2 Cascadia megathrust geometry \cite{Hayes_2018_Slab2ComprehensiveSubduction}, whereas recent regional-scale seismic imaging of the Cascadia margin \cite{Carbotte_2024_SubductingPlateStructure} suggests locally different, and in places shallower, subduction interface. Likewise, although a Nankai-style megasplay fault system is not likely to exist in Cascadia \cite{Lucas_2023_NoEvidenceActive}, smaller regional splays may contribute to source complexity and tsunami generation \cite<e.g.,>{Wendt_2009_TsunamisSplayFault,VanZelst_2022_EarthquakeRuptureMultiple, Biemiller_2025_StructuralControlsSplay}. Future work may include fully-coupled simulations with widespread off-fault plastic deformation  \cite{Ulrich_2022_StressRigiditySediment,Ma_2023_WedgePlasticityMinimalist, Ma_2025_WedgeInelasticityFully}, and application to other subduction margins \cite{Yao_2020_RuptureDynamics2012, Wirp_2025_HellenicArcTsunami, Wong_2026_DynamicRestrengtheningFault}.

\subsection{Implications for real-time tsunami forecasting}

Given real-time near-field pressure transients, our Bayesian inversion-based framework infers seafloor motion and produces probabilistic tsunami wave height forecasts 
at selected offshore target locations within fractions of a second (Supporting Information, Table~S3) \cite{henneking2025bell}.
If only tsunami forecasts are required, without reconstruction of the full seafloor displacement field, the real-time inference can be performed with minimal computing infrastructure, for example, on a laptop (Supporting Information, Table~S3) \cite{henneking2026goal}.
Here, we demonstrate the framework for 21 forecast locations. Noting that even a limited number of tsunami wave height forecasts is informative for early warning purposes \cite{Liu_2021_ComparisonMachineLearning}, this number could readily be increased to several hundred forecast locations while retaining real-time forecast capability with modest compute resources.

To evaluate the acoustic--gravity Bayesian inversion framework, we use complex, earthquake-induced seafloor velocity fields generated by fully-coupled dynamic rupture simulations with elastic--acoustic--gravity wave propagation.
One could, in principle, solve instead an inverse problem for the fault slip using the coupled Earth--ocean system.
The 3-field elastic--acoustic--gravity system remains linear time-invariant, thus, our current inversion framework might still be executable in real time. However, in addition to much greater offline cost due to solving the computationally more demanding 3-field system (Supplementary Information, Text S1), there are several reasons to invert for seafloor motion using the acoustic--gravity model over using the fully-coupled system in the inversion.

Inverting for fault slip rather than seafloor motion would introduce additional uncertainty from seismic velocity structure and fault geometry into the inverse problem \cite<e.g.,>{Beresnev_2003_UncertaintiesFiniteFaultSlip, Duputel_2014_AccountingPredictionUncertainty, Ragon_2018_AccountingUncertainFault}. 
By contrast, uncertainties in ocean acoustic wave speed due to seawater temperature, salinity, bulk modulus, and density are comparatively small \cite<e.g.,>{Mackenzie_1981_NinetermEquationSound, Dushaw_1993_EquationsSpeedSound}. 
In addition, excluding the elastic solid in the inversion avoids assuming that seafloor uplift arises only from coseismic elastic deformation, and in principle allows extension to other tsunami sources, such as mass slumping, landslides, (volcanic) explosions, or 
inelastic coseismic deformation ~\cite<e.g.,>{Duffy1992, Grilli2005, Pedersen2015, Schulten2019, Schindele2024, Svennevig2024, Du_2025_WedgeInelasticityFully}.
Because our goal is to infer seafloor motion in real time and use it to forecast tsunami propagation, it is not necessary to include the elastic solid in the inverse problem: although seafloor motion in our models originates from the coupled solid Earth--ocean system, 
the mapping from seafloor motion to the induced pressure transients and gravity-wave excitation is captured by the acoustic–gravity model. 

Our results indicate that relatively sparse near-field pressure observations are sufficient to infer the smooth, tsunami-relevant components of seafloor motion and that these components account well for the surface gravity wave generation.
This is reflected in the substantially smaller relative errors of the tsunami wave height forecasts (18.1\%--22.1\%) compared to those of the inferred seafloor displacements (27.7\%--36.1\%).
Furthermore, the inference of these smooth components deteriorates only slowly as the number of sensors is reduced from 600 to 175, with relative errors in the inferred seafloor displacement field increasing from 29.6\% to 36.1\% for the margin-wide rupture and from 27.7\% to 31.7\% for the partial rupture scenarios. The corresponding relative errors of the tsunami wave height forecasts increase even more slowly, from 18.6\% to 22.1\% and from 18.1\% to 19.6\%, respectively.

The modest degradation in tsunami forecast accuracy with decreasing sensors from 600 to 175 indicates that the forecasts are relatively robust to sensor sparsification. This may suggest that further reducing the number of sensors would still achieve reasonably accurate forecasts across the CSZ, although the inferred seafloor motion and associated uncertainties deteriorate more strongly, as quantified by the Bayesian framework through the relative increase of the credible intervals from the left column to the right column plots of \Cref{fig:forecast-results}.

We note that the 175-sensor network considered here was placed ad-hoc, without optimization of sensor locations.
Rather than prescribing sensor locations a priori, future work can address the design challenge of optimal sensor placement, given a fixed budget.
In the Bayesian framework, future work can pose such an \emph{optimal experimental design} (OED) problem \cite<e.g.,>{atkinson2007oed, pukelsheim2006oed} as an optimization problem that maximizes the expected information gain from the sensor observations, resulting in the smallest uncertainties of the inferred solution \cite<e.g.,>{ghattas2021learning, alexanderian2016oed, venkat2026oed}.
This information gain could also be tailored to be goal-oriented \cite<e.g.,>{attia2018goal, wu2023goal}, prioritizing sensor locations that most reduce tsunami forecast uncertainty in location and time most relevant for tsunami early warning and hazard or risk mitigation in Cascadia.

%% file: sections/conclusions.tex
This study applies a physics-based Bayesian inversion framework for real-time, data-driven tsunami forecasting in Cascadia that can accommodate complex source behavior. 
Following an earthquake, the framework assimilates near-field ocean-bottom pressure observations to infer the earthquake-induced spatiotemporal seafloor motion, and uses this inverse solution to forecast tsunami propagation.
The inversion and tsunami forecasting employ an acoustic--gravity model that captures coupled ocean acoustic and tsunami wave propagation.
Both the inferred seafloor motion and the tsunami forecasts are accompanied by quantified uncertainty estimates.
Because the most expensive computations of the inversion are shifted to an offline precomputation stage performed before any earthquake occurs, real-time inference can be carried out in less than a second once observations are available.

Using two fully coupled Cascadia dynamic rupture–tsunami scenarios, we show that early tsunami-source discrimination is nontrivial, yet possible on the timescale of minutes. The initial near-field ocean-bottom pressure signals are similar for partial and margin-wide ruptures. However, on the timescale of minutes after earthquake rupture onset, the scenarios become distinguishable because high-amplitude rupture-driven vertical sea surface velocity signals decay after the southward partial rupture arrests near 46°N, whereas the margin-wide rupture continues southward and arrests near 42°N. 
The northward rupture fronts in both scenarios terminate abruptly at the northern end of the megathrust and radiate stopping phases with reversed polarity. 
Together with back-propagating rupture fronts, these complex dynamic rupture source effects imprint distinct signals on the sea surface wavefield. 

For a hypothetical offshore pressure network of 175 sensors, comparable in scale to Japan’s S-net, the resulting relative tsunami forecast errors are 22.1\% for the margin-wide rupture and 19.6\% for the partial rupture, representing only a modest degradation relative to a denser 600-sensor configuration.
Together, these results are a step toward physics-consistent, real-time probabilistic tsunami forecasting in the Cascadia Subduction Zone and motivate future offshore ocean-bottom sensor deployments.

%% file: sections/supporting.tex
\makeatletter
\long\def\@makecaption#1#2{%
  \vskip\abovecaptionskip
  \begin{center}#2\end{center}%
  \vskip\belowcaptionskip}
\makeatother

\noindent\textbf{Text S1. 3D fully-coupled earthquake dynamic rupture–tsunami simulations}

To model both 3D fully-coupled earthquake dynamic rupture and tsunami scenarios used in this study, we use SeisSol which is based on the Arbitrary high-order DERivative-Discontinuous Galerkin (ADER-DG) method \cite{Kaser_2006_ArbitraryHighorderDiscontinuous,Dumbser_2006_ArbitraryHighorderDiscontinuous}.
SeisSol has been verified against a set of community benchmark problems, including a recent 3D fully-coupled setup  \cite<e.g.,>{Pelties_2014_VerificationADERDGMethod, Harris_2018_SuiteExercisesVerifying, Kutschera_2025_CRESCENTSCECUSGS}, and has been optimized for high-performance computing \cite<e.g.,>{Uphoff_2017_ExtremeScaleMultiPhysics}.
We use SeisSol with sixth-order accuracy in time and space (i.e., polynomial order of $p=5$) and simulate 420~s of earthquake–tsunami generation. Both fully-coupled simulations were run on the \emph{Frontera} supercomputer \cite{Stanzione_2020_FronteraEvolutionLeadership} using 102 nodes (5,712 cores), each requiring about 175~kCPUh.

The fully-coupled simulations required a modification to the unstructured tetrahedral mesh of \citeA{Glehman_2025_PartialRupturesGoverned}, incorporating the ocean acoustic water layer atop the elastic domain (solid Earth). Thus, we rebuild the 3D unstructured, tetrahedral mesh with the same 1.5~km on-fault mesh resolution \cite{Glehman_2025_PartialRupturesGoverned} and a resolution of 0.4~km in the acoustic medium. The resulting mesh is composed of a total of 103,386,944 elements. Out of those, 43,633,138 elements represent the solid Earth and 59,753,806 elements the water layer.

We note that the modeling domain for the ocean acoustic water layer in our fully-coupled simulations is spatially limited and involves sharp bends due to the high computational cost and the non-trivial intersection of the water layer with the coastal bathymetry \cite{Wirp_2025_HellenicArcTsunami}. 
Thus, along the edges, the faster-propagating seismo-acoustic waves may be reflected in the E–W direction, as the width of the water layer is shorter compared to its length. 
However, the amplitudes of those boundary reflections are $\sim$80\% smaller compared to the physically generated seismo-acoustic waves and $\sim$99\% smaller than the peak tsunami amplitude (Supporting Information, Movie~S3).

Properties of the margin-wide and partial dynamic rupture scenario are listed in Supporting Information, Table S1. This expands on the reported rupture duration, peak slip rate, ruptured area, estimated stress drop, and the peak vertical ground displacements of the dynamic rupture scenarios given in \citeA{Glehman_2025_PartialRupturesGoverned}.

The modeled seafloor velocities %
are sampled at 1~Hz in SeisSol before being used in the Bayesian inversion framework. A higher temporal sampling as well as an increased simulation duration beyond 420~s would result in a larger inverse problem (i.e., larger parameter space), detailed in Text~S2.

\noindent\textbf{Text S2. 3D acoustic--gravity wave propagation simulations}

The acoustic--gravity PDEs, as well as the corresponding adjoint PDEs, are discretized using high-order finite element methods implemented in the MFEM library~\cite{anderson2021mfem}. 
Specifically, for both the forward and the adjoint problem, we solve a mixed variational formulation that is discretized with the Galerkin finite element method and explicit fourth-order Runge--Kutta (RK4) time-stepping. The finite element discretization uses fourth-order continuous ($H^1$-conforming) scalar-valued pressure and third-order discontinuous ($L^2$-conforming) velocity components.
The computational domain is meshed using hexahedral elements, with a spatial resolution of 300~meters, resulting in a state vector of $\sim$3.74 billion degrees of freedom. The RK4 timestep size is dictated by the Courant--Friedrichs--Lewy (CFL) stability condition. For a simulation time of 420~s, a total of 336,000 RK4 timesteps are performed. 

The simulation time per PDE solution is 52~minutes on 512 A100 GPUs on NERSC's \emph{Perlmutter} supercomputer. The seafloor velocity parameter space is discretized with linear elements over the seafloor boundary of the 3D mesh, with $\numparam=$~2,416,530 spatial grid points across $\numtime=$~420 temporal snapshots, yielding a total parameter space dimension of $\sim$1.015 billion. The pressure observables at the 175 sensor locations are stored at 420 time instances (one per second), for a total data dimension 73,500. This 1~Hz sampling of pressure data (capturing wave frequencies up to the Nyquist frequency of 0.5~Hz) was chosen corresponding to the 1~Hz sampling rate of the input seafloor velocity parameter field generated by the fully-coupled earthquake–tsunami simulations (see Text~S1).

In~\citeA{henneking2025bell}, the major computational cost of the inverse solution arose from precomputing one solution to the adjoint acoustic--gravity PDEs for each of the 600 sensor locations. The adjoint PDE solutions of the 175-sensor configuration used in the present work is obtained by subselecting from the original 600-sensor setup. Because of this subselecting, the computational cost of constructing the 175-sensor configuration reduces to the much cheaper computations with the efficient FFT-based solver (see Table S2).

\begin{table}[hp]
    \centering
    \begin{tabular}{l | c | c}
        \toprule
                                                       & Partial rupture & Margin-wide rupture \\
        \midrule\midrule
        \citeA{Glehman_2025_PartialRupturesGoverned}   & Model 2         & Model 16            \\
        \midrule
        Static friction coefficient ($\mu_s$)          & 0.6             & 0.6                 \\
        Dynamic friction coefficient ($\mu_d$)         & 0.1             & 0.1                 \\
        Critical slip-weakening distance ($D_c$) [m]   & 1.0             & 1.0                 \\
        Nucleation radius ($r_{\text{crit}}$) [km]     & 3400            & 4400                \\
        Nucleation depth [km]                          & 16              & 16                  \\
        Pore fluid ratio ($\gamma$)                    & 0.97            & 0.97                \\
        \midrule
        Moment magnitude $M_W$                         & 8.6             & 8.7                 \\
        Rupture duration [s]                           & $\sim$140       & $\sim$270           \\
        Peak slip rate [m\,s$^{-1}$]                   & 2.26            & 2.37                \\
        Peak slip [m]                                  & 11.25           & 11.34               \\
        Average slip [m]$^{*}$                         & $\sim$4.9       & $\sim$4.8           \\
        Average rupture velocity [km\,s$^{-1}$]$^{*}$  & $\sim$2.14      & $\sim$2.12          \\
        Ruptured area [km$^2$]$^{*}$                   & 58,349.6        & 86,767.3            \\
        Stress drop [MPa]$^{*}$                        & 1.33            & 1.45                \\
        Peak seafloor uplift (final) [m]               & 1.59            & 1.69                \\
        Peak seafloor subsidence (final) [m]           & -2.03           & -2.07               \\ 
        \bottomrule
    \end{tabular}
    \caption{\textbf{Table S1.} Dynamic rupture parameters and key results for the two fully-coupled dynamic rupture scenarios expanded from \citeA{Glehman_2025_PartialRupturesGoverned}. $^{*}$Considering those parts of the fault where the slip exceeds 0.2~m.}
\end{table}

\begin{table}[hp]
    \centering
    \begin{tabular}{@{}llll@{}}
    \toprule
    Symbol & Description & Value & Unit \\
    \midrule
    $g$ & Gravity & 9.807 & m\,s$^{-2}$ \\
    $\rho$ & Density of seawater & $1.025 \cdot 10^3$ & kg\,m$^{-3}$ \\
    $K$ & Bulk modulus of seawater & $2.340 \cdot 10^9$ & Pa \\
    $c$ & Sound speed in seawater & $1.511 \cdot 10^3$ & m\,s$^{-1}$ \\
    $Z$ & Specific acoustic impedance of seawater & $1.549 \cdot 10^6$ & kg\,m$^{-2}$\,s$^{-1}$ \\
    \bottomrule
    \end{tabular}
    \caption{\textbf{Table S2.} Physical parameters used in the acoustic--gravity model, specified to four significant digits of accuracy.}
\end{table}

\begin{table}[hp]
    \centering
    \addtolength{\tabcolsep}{-1pt}
\addtolength{\tabcolsep}{-1pt}
\begin{tabular}{@{}rl@{\hspace{4pt}}rll@{}}
    \toprule
    Phase & Task & GPUs & Time (600 sensors) & Time (175-sensor subset) \\
    \midrule
    1 &form $\ptomat: \paramvec \mapsto \datavec$ & 512&
    600 $\!\times\!$ 52~m $\sim$ 520~h &
    - \\
    & form $\ptqmat: \paramvec \mapsto \qoivec$ &512&
    21 $\!\times\!$ 52~m $\sim$ 18~h &
    - \\
    2  &form $\Gmat^* \coloneqq \bfpriorcov \ptomat^*$ &16&
    600 $\!\times\!$ 4.5~s $\sim$ 45~m &
    - \\
    & form $\Gmat_q^* \coloneqq \bfpriorcov \ptqmat^*$ &16&
    21 $\!\times\!$ 4.5~s  $\sim$ 1.5~m &
    - \\
    &form $\Kmat \coloneqq \bfnoise + \ptomat \Gmat^*$ &512&
    252k $\!\times\!$ 24~ms  $\sim$ 100~m &
    - \\
    & factorize $\Kmat$ &25& 22~s &
    4~s$^{\ast}$ \\
    3 &compute $\bfpostcovq$ &512& 8,820 $\!\times\!$ 150~ms $\sim$ 25~m &
    8,820 $\!\times\!$ 100~ms $\sim$ 15~m$^{\dag}$ \\
    & compute $\Qmat : \datavec \mapsto \qoivec$ &512& 8,820 $\!\times\!$ 100~ms $\sim$ 15~m &
    8,820 $\!\times\!$ 100~ms $\sim$ 15~m$^{\dag}$ \\
    4 & infer parameters $\bfmmap$ &512& $<$ 0.2~s &
    $<$ 0.1~s$^{\dag}$ \\
    & predict QoI $\bfqmap$ &1& $<$ 1~ms &
    $<$ 1~ms \\
    \bottomrule
    \multicolumn{5}{@{}l@{}}{\footnotesize $^{\ast}$Computed using 4 GPUs \quad $^{\dag}$Computed using 128 GPUs}
\end{tabular}
    \caption{\textbf{Table S3.} Compute time for each phase of the Bayesian inversion framework performed on Perlmutter A100 GPU nodes. Setting up a 175-sensor network sub-selected from the 600 sensor locations computed in \citeA{henneking2025bell} can be done at a relatively low computational cost. The online inference (Phase~4) of seafloor velocity parameters from pressure observations is computed in less than 0.2~seconds on 512 GPUs. The tsunami wave heights at 21 forecasting locations can be obtained on a single GPU (or a laptop) in 1~ms. The explicit reconstruction of the seafloor velocity parameter field is not required to compute the tsunami forecasts. That is, if only tsunami forecasts are needed, the entire online phase of tsunami forecasting from seafloor pressure data can be performed in real time on a single GPU (or a laptop).}
\end{table}

\clearpage
\begin{figure}[hp]
    \centering
    \includegraphics[width=0.9\textwidth]
    {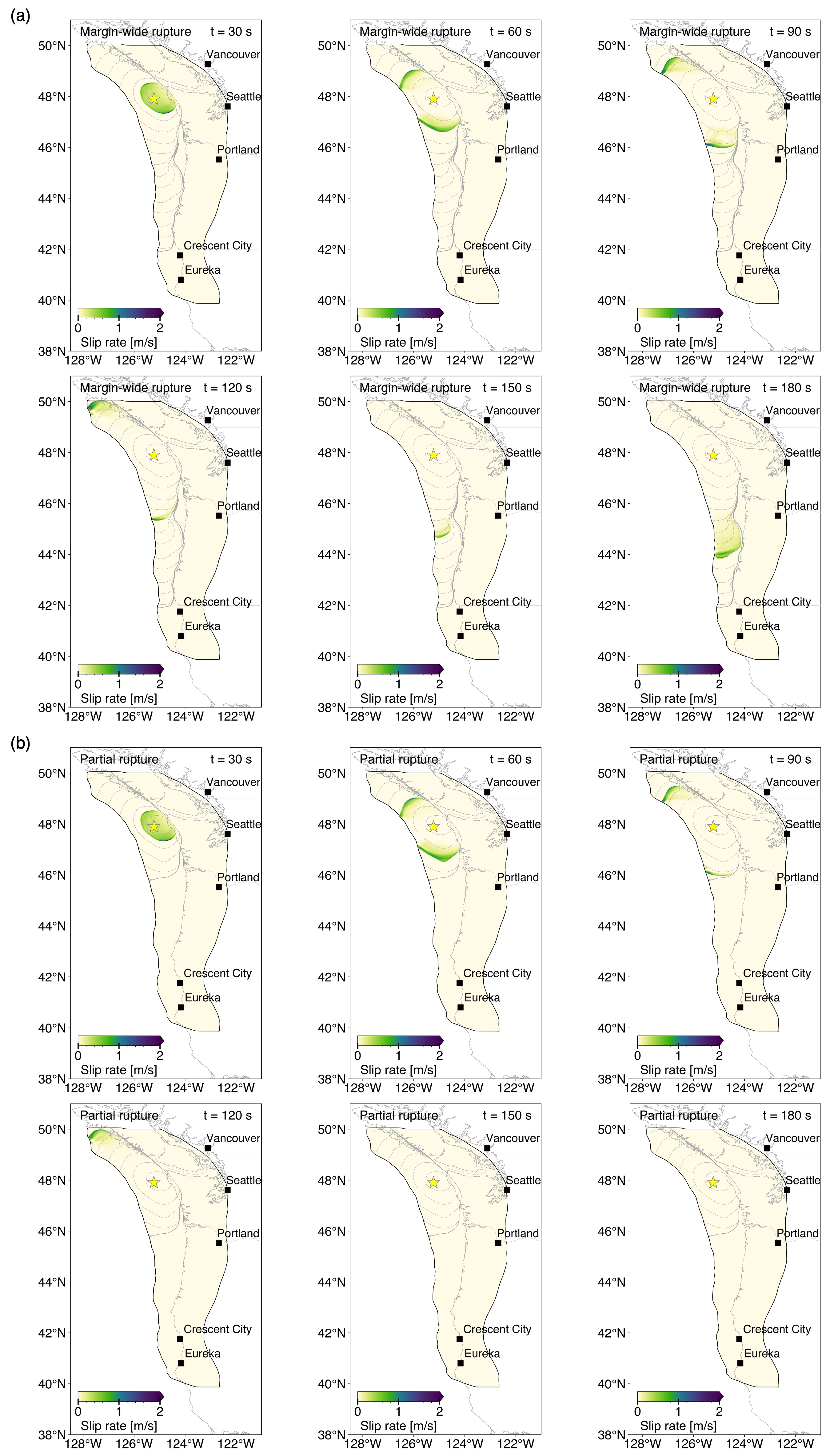}
    \caption{\textbf{Figure S1.} Snapshots of the slip rate every 30~s for (a) the margin-wide rupture and (b) the partial rupture scenario. The yellow star marks the hypocenter. Note the earlier rupture arrest for the partial rupture scenario compared to the margin-wide rupture. Animations of the slip rate are uploaded as additional, supplementary movie.}
\end{figure}

\begin{figure}[hp]
    \centering
    \includegraphics[width=\textwidth]
    {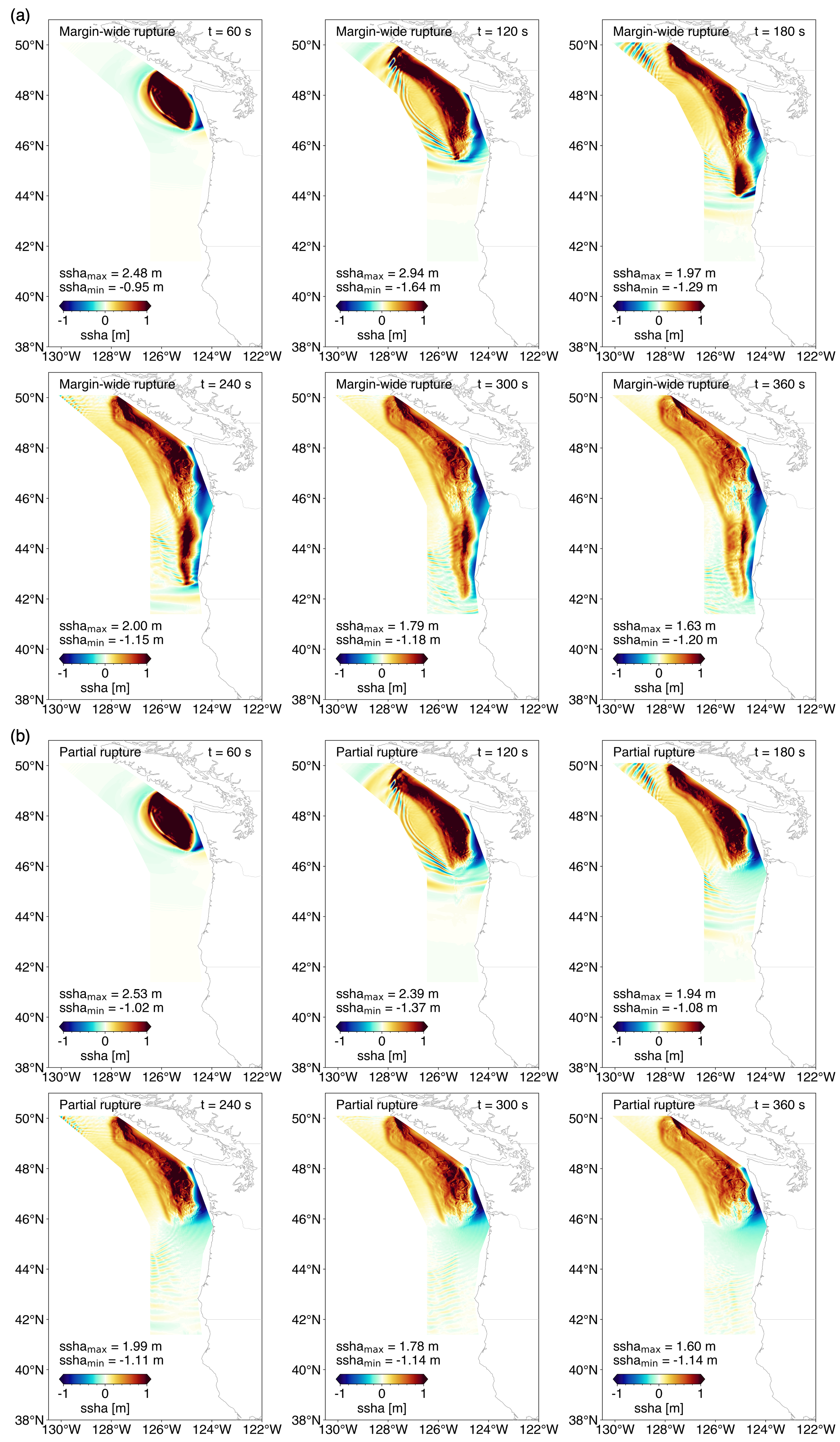}
    \caption{\textbf{Figure S2.} Snapshots of the seismo-acoustic and gravity wave (i.e., tsunami) excitation at the sea surface as indicated by the sea surface height anomaly (ssha) every 60~s for (a) the margin-wide rupture and (b) the partial rupture scenario. Animations of the time-dependent ssha are uploaded as additional, supplementary movie.}
\end{figure}

\clearpage

\begin{figure}[htp] 
    \centering
    \includegraphics[width=\textwidth]
    {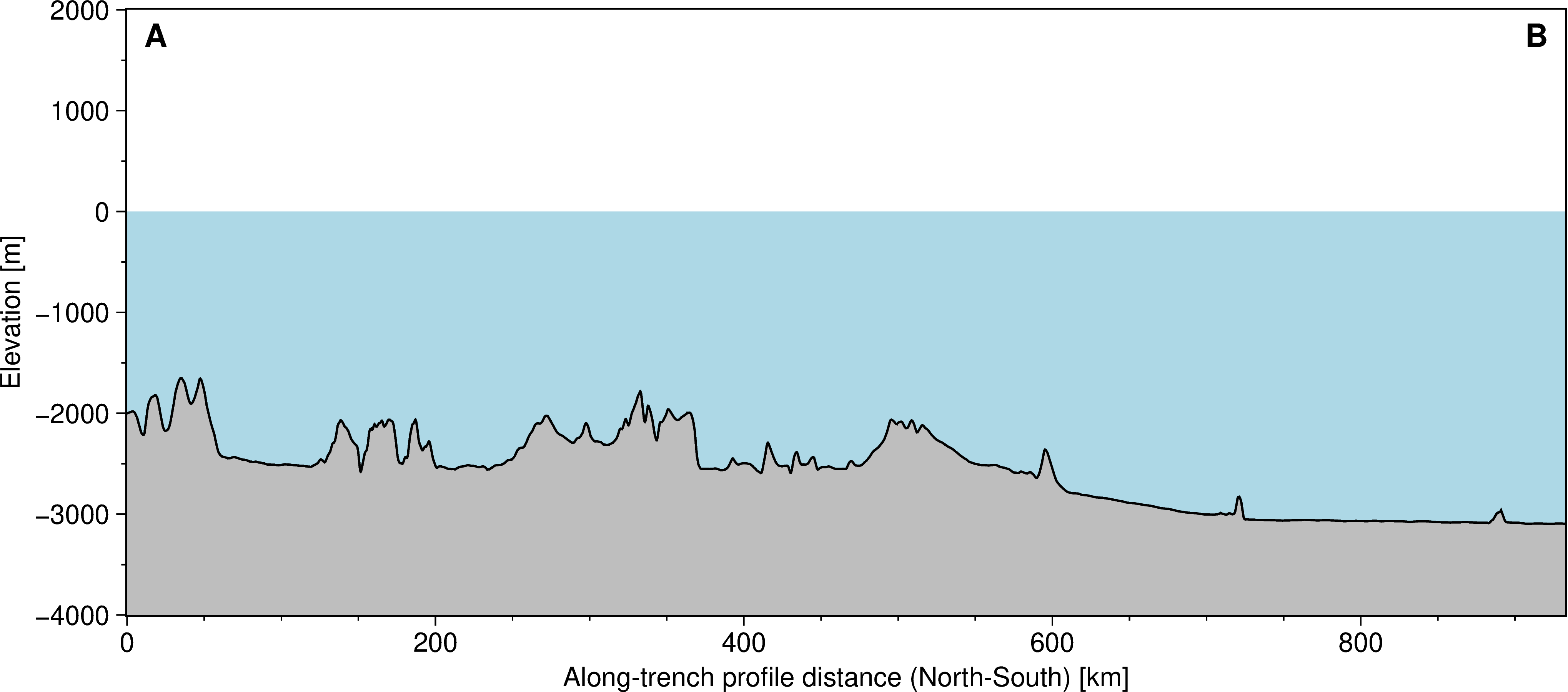}
    \caption{\textbf{Figure S3.} Enlarged seafloor elevation along the trench profile from North to South (A--B) from Figure~\ref{fig:figure01}a.}
    
    \vspace{1em} 

    \centering
    \includegraphics[width=\textwidth]
    {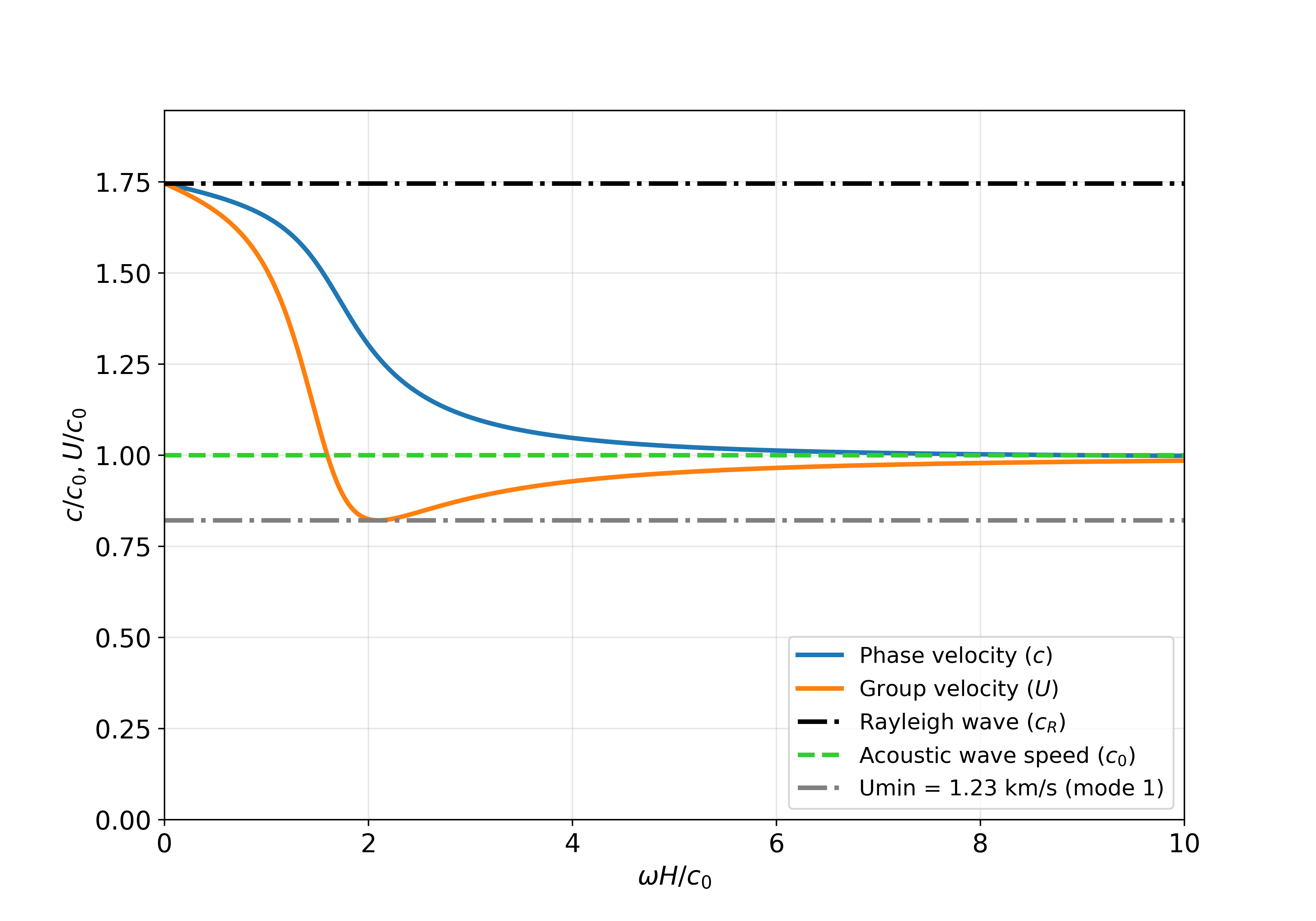}
    \caption{\textbf{Figure S4.} Phase ($c$) and group velocity ($U$), normalized by the ocean acoustic wave speed ($c_0$) for the mode 1 ($n=1$) oceanic Rayleigh wave, plotted as a function of dimensionless angular frequency ($\omega H/c_0$) following \citeA{Abrahams_2023_ComparisonMethodsCoupled}. Here, we assume an average water depth of $H=2554$~m from the along-trench profile in Figure~S3. Given that the hypocenters of the margin-wide and partial rupture scenario are both at 16~km depth, we here assume a by thickness-weighted average for the material parameters of the used 1D subsurface structure \cite{Sallares_2019_UpperplateRigidityDetermines} for the uppermost 16~km to compute the solid Earth Rayleigh wave velocity ($c_R$) and the mode 1 minimum group velocity ($U_{\min}$).}
\end{figure}

\clearpage

\begin{figure}[htp] 
    \centering
    \includegraphics[width=\textwidth]
    {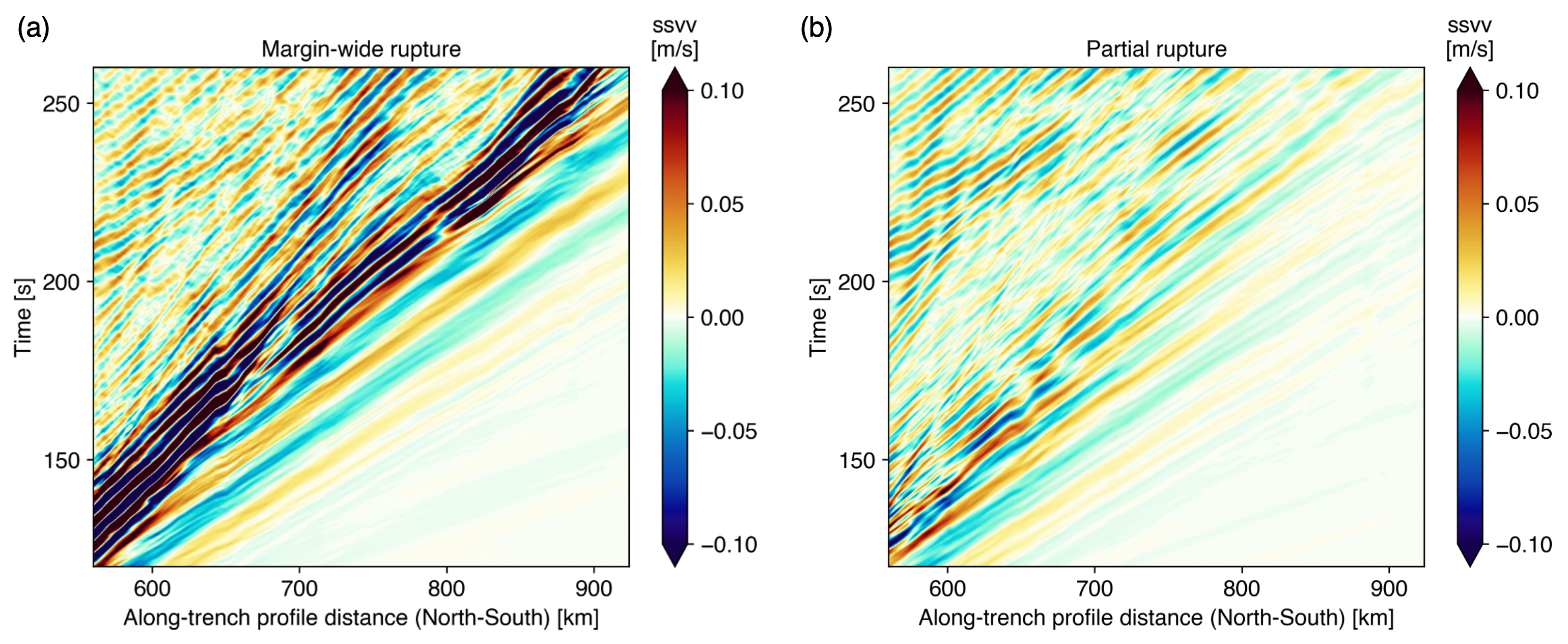}
    \caption{\textbf{Figure S5.} Zoom into the space-time evolution of the sea surface vertical velocities (ssvv) for (a) the margin-wide rupture and (b) the partial rupture scenario, showing the amplitude-decaying oceanic Rayleigh waves once the partial rupture has arrested in comparison to the ongoing margin-wide rupture scenario.}
\end{figure}

\noindent\textbf{Movie S1.}
Slip rate animations for the margin-wide rupture (left) and the partial rupture (right) scenario. The yellow star marks the hypocenter.

\noindent\textbf{Movie S2.}
Sea surface height anomaly (ssha) animations including the seismo-acoustic and gravity wave (i.e., tsunami) excitation for the margin-wide rupture (left) and the partial rupture (right) scenario.

\noindent\textbf{Movie S3.}
Animations of the vertical particle velocity ($v_3$) at the sea surface (upper row; referred to as sea surface vertical velocities (ssvv) in the main text) and the seafloor (bottom row) for the margin-wide and partial rupture scenario. Note the different extent for the fully-coupled modeling domain (see Figure~\ref{fig:figure01})  versus the entire solid Earth modeling domain. 

\noindent\textbf{Movie S4.}
True (left) and inferred (middle and right) seafloor normal displacement fields over the 420-second simulation time for the margin-wide rupture scenario. The ``ground truth'' on the left was computed with the fully-coupled dynamic rupture and tsunami forward model. The inferred displacements were obtained from the acoustic–gravity model inversion using synthetic ocean bottom pressure data from the hypothesized 600-sensor network (middle) and 175-sensor network (right).

\noindent\textbf{Movie S5.}
True (left) and inferred (middle and right) seafloor normal displacement fields over the 420-second simulation time for the partial rupture scenario. The ``ground truth'' on the left was computed with the fully-coupled dynamic rupture and tsunami forward model. The inferred displacements were obtained from the acoustic–gravity model inversion using synthetic ocean bottom pressure data from the hypothesized 600-sensor network (middle) and 175-sensor network (right).

%% file: main.bib
@string{ACTANUM="Acta Numerica"}

@string{CAMWA="Computers \& Mathematics with Applications"}

@string{CGEO="Computational Geosciences"}

@string{JCP="Journal of Computational Physics"}

@string{JGRSE="Journal of Geophysical Research: Solid Earth"}

@string{JSC="SIAM Journal on Scientific Computing"}

@string{PAGEO="Pure and Applied Geophysics"}

@article{Maeda2015,
  title = {Successive Estimation of a Tsunami Wavefield without Earthquake Source Data: {{A}} Data Assimilation Approach toward Real-Time Tsunami Forecasting},
  author = {Maeda, Takuto and Obara, Kazushige and Shinohara, Masanao and Kanazawa, Toshihiko and Uehira, Kenji},
  year = {2015},
  journal = {Geophysical Research Letters},
  volume = {42},
  number = {19},
  pages = {7923--7932},
  issn = {1944-8007},
  doi = {10.1002/2015GL065588}
}

@article{Gusman2016,
  title = {Tsunami Data Assimilation of {{Cascadia}} Seafloor Pressure Gauge Records from the 2012 {{Haida Gwaii}} Earthquake},
  author = {Gusman, Aditya Riadi and Sheehan, Anne F. and Satake, Kenji and Heidarzadeh, Mohammad and Mulia, Iyan Eka and Maeda, Takuto},
  year = {2016},
  journal = {Geophysical Research Letters},
  volume = {43},
  number = {9},
  pages = {4189--4196},
  issn = {1944-8007},
  doi = {10.1002/2016GL068368}
}

@article{Pedersen2015,
  title={On the characteristics of landslide tsunamis},
  author={Løvholt, F. and Pedersen, G. and Harbitz, C. B. and Glimsdal, S. and Kim, J.},
  journal={Philosophical Transactions of the Royal Society A: Mathematical, Physical and Engineering Sciences},
  volume={373},
  number={2053},
  pages = {20140376},
  year={2015},
  publisher={The Royal Society},
  doi = {10.1098/rsta.2014.0376}
}

@article{Duffy1992,
  title={On the generation of oceanic surface waves by underwater volcanic explosions},
  author={Duffy, Dean G.},
  journal={Journal of volcanology and geothermal research},
  volume={50},
  number={3},
  pages={323--344},
  year={1992},
  publisher={Elsevier}
}

@article{Schulten2019,
    author = {Schulten, Irena and Mosher, David C. and Piper, David J. W. and Krastel, Sebastian},
    title = {A Massive Slump on the {St. Pierre Slope}, A New Perspective on the 1929 {Grand Banks} Submarine Landslide},
    journal = {Journal of Geophysical Research: Solid Earth},
    volume = {124},
    number = {8},
    pages = {7538-7561},
    doi = {10.1029/2018JB017066},
    year = {2019}
}

@article{Schindele2024,
  title={A review of tsunamis generated by volcanoes ({TGV}) source mechanism, modelling, monitoring and warning systems},
  author={Schindel{\'e}, Fran{\c{c}}ois and Kong, Laura and Lane, Emily M and Paris, Rapha{\"e}l and Ripepe, Maurizio and Titov, Vasily and Bailey, Rick},
  journal={Pure and Applied Geophysics},
  volume={181},
  number={6},
  pages={1745--1792},
  year={2024},
  publisher={Springer}
}

@article{Grilli2005,
  title={Tsunami generation by submarine mass failure. {I}: {Modeling}, experimental validation, and sensitivity analyses},
  author={Grilli, St{\'e}phan T and Watts, Philip},
  journal={Journal of waterway, port, coastal, and ocean engineering},
  volume={131},
  number={6},
  pages={283--297},
  year={2005},
  publisher={American Society of Civil Engineers}
}

@article{Svennevig2024,
    author = {Kristian Svennevig  and Stephen P. Hicks  and Thomas Forbriger  and Thomas Lecocq  and Rudolf Widmer-Schnidrig  and Anne Mangeney  and Clément Hibert  and Niels J. Korsgaard  and Antoine Lucas  and Claudio Satriano  and Robert E. Anthony  and Aurélien Mordret  and Sven Schippkus  and Søren Rysgaard  and Wieter Boone  and Steven J. Gibbons  and Kristen L. Cook  and Sylfest Glimsdal  and Finn Løvholt  and Koen Van Noten  and Jelle D. Assink  and Alexis Marboeuf  and Anthony Lomax  and Kris Vanneste  and Taka’aki Taira  and Matteo Spagnolo  and Raphael De Plaen  and Paula Koelemeijer  and Carl Ebeling  and Andrea Cannata  and William D. Harcourt  and David G. Cornwell  and Corentin Caudron  and Piero Poli  and Pascal Bernard  and Eric Larose  and Eleonore Stutzmann  and Peter H. Voss  and Bjorn Lund  and Flavio Cannavo  and Manuel J. Castro-Díaz  and Esteban Chaves  and Trine Dahl-Jensen  and Nicolas De Pinho Dias  and Aline Déprez  and Roeland Develter  and Douglas Dreger  and Läslo G. Evers  and Enrique D. Fernández-Nieto  and Ana M. G. Ferreira  and Gareth Funning  and Alice-Agnes Gabriel  and Marc Hendrickx  and Alan L. Kafka  and Marie Keiding  and Jeffrey Kerby  and Shfaqat A. Khan  and Andreas Kjær Dideriksen  and Oliver D. Lamb  and Tine B. Larsen  and Bradley Lipovsky  and Ikha Magdalena  and Jean-Philippe Malet  and Mikkel Myrup  and Luis Rivera  and Eugenio Ruiz-Castillo  and Selina Wetter  and Bastien Wirtz },
    title = {A rockslide-generated tsunami in a {Greenland} fjord rang {Earth} for 9 days},
    journal = {Science},
    volume = {385},
    number = {6714},
    pages = {1196--1205},
    year = {2024},
    doi = {10.1126/science.adm9247}
}

@article{Kubota_2021_ExtractingNearFieldSeismograms,
  title = {Extracting near-field seismograms From ocean-bottom pressure gauge inside the focal area: {Application} to the 2011 {Mw} 9.1 {Tohoku-Oki} earthquake},
  author = {Kubota, Tatsuya and Saito, Tatsuhiko and Tsushima, Hiroaki and Hino, Ryota and Ohta, Yusaku and Suzuki, Syuichi and Inazu, Daisuke},
  year = 2021,
  journal = {Geophysical Research Letters},
  volume = {48},
  number = {7},
  pages = {e2020GL091664},
  issn = {1944-8007},
  doi = {10.1029/2020GL091664}
}

@article{Ma_2023_WedgePlasticityMinimalist,
  title = {Wedge Plasticity and a Minimalist Dynamic Rupture Model for the 2011 {{MW}} 9.1 {{Tohoku-Oki}} Earthquake and Tsunami},
  author = {Ma, Shuo},
  year = 2023,
  journal = {Tectonophysics},
  volume = {869},
  pages = {230146},
  issn = {0040-1951},
  doi = {10.1016/j.tecto.2023.230146}
}

@article{Ulrich_2022_StressRigiditySediment,
  title = {Stress, Rigidity and Sediment Strength Control Megathrust Earthquake and Tsunami Dynamics},
  author = {Ulrich, Thomas and Gabriel, Alice Agnes and Madden, Elizabeth H.},
  year = 2022,
  journal = {Nature Geoscience},
  volume = {15},
  number = {1},
  pages = {67--73},
  issn = {1752-0908},
  doi = {10.1038/s41561-021-00863-5}
}

@article{Wong_2026_DynamicRestrengtheningFault,
  title = {Dynamic restrengthening and fault heterogeneity explain megathrust earthquake complexity},
  author = {Wong, Jeremy Wing Ching and Gabriel, Alice-Agnes and Fan, Wenyuan},
  year = {2026},
  journal = {arXiv:2505.08973},
  doi = {10.48550/arXiv.2505.08973}
}

@article{VanZelst_2022_EarthquakeRuptureMultiple,
  title = {Earthquake rupture on multiple splay faults and its effect on tsunamis},
  author = {{van Zelst}, I. and Rannabauer, L. and Gabriel, A. A. and {van Dinther}, Y.},
  year = 2022,
  journal = {Journal of Geophysical Research: Solid Earth},
  volume = {127},
  number = {8},
  pages = {e2022JB024300},
  issn = {2169-9356},
  doi = {10.1029/2022JB024300}
}

@article{Melgar_2021_WasJanuary26th,
  title = {Was the {January} 26th, 1700 {Cascadia} earthquake part of a rupture sequence?},
  author = {Melgar, Diego},
  year = {2021},
  journal = {Journal of Geophysical Research: Solid Earth},
  volume = {126},
  number = {10},
  pages = {e2021JB021822},
  doi = {10.1029/2021JB021822}
}

@Article{Tsushima2009,
    author = {Tsushima, Hiroaki and Hino, Ryota and Fujimoto, Hiromi and Tanioka, Yuichiro and Imamura, Fumihiko},
    title = {Near-field tsunami forecasting from cabled ocean bottom pressure data},
    journal = JGRSE,
    volume = {114},
    number = {B6},
    pages = {},
    doi = {10.1029/2008JB005988},
    year = {2009}}

@Inproceedings{leveque2018cascadia,
   title={Developing a warning system for inbound tsunamis from the {Cascadia} subduction zone},
   author={LeVeque, Randall J. and Bodin, Paul and Cram, Geoffrey and Crowell, Brendan W. and González, Frank I. and Harrington, Michael and Manalang, Dana and Melgar, Diego and Schmidt, David A. and Vidale, John E. and Vogl, Christopher J. and Wilcock, William S. D.},
   booktitle={OCEANS 2018 MTS/IEEE Charleston},
   pages={1--10},
   year={2018},
   organization={IEEE},
   doi={10.1109/OCEANS.2018.8604709}}

@Techreport{schmidt2019monitoring,
   title={Earthquake and tsunami early warning on the {Cascadia} subduction zone: {A} feasibility study for an offshore geophysical monitoring network},
   author={Schmidt, David and Wilcock, William and LeVeque, Randall J. and Gonzalez, Frank and Cram, Geoffrey and Manalang, Dana and Harrington, Mike and Roland, Emily and Bodin, Paul},
   institution={University of Washington},
   year={2019},
   note={\url{https://hdl.handle.net/1773/50968}}}

@Article{anderson2021mfem,
   title = {{MFEM: A modular finite element methods library}},
   author = {Anderson, Robert and Andrej, Julian and Barker, Andrew and Bramwell, Jamie and Camier, Jean-Sylvain and Cerveny, Jakub and Dobrev, Veselin and Dudouit, Yohann and Fisher, Aaron and Kolev, Tzanio and Pazner, Will and Stowell, Mark and Tomov, Vladimir and Akkerman, Ido and Dahm, Johann and Medina, David and Zampini, Stefano},
   journal = CAMWA,
   volume = {81},
   pages = {42--74},
   year = {2021},
   publisher = {Elsevier},
   doi = {10.1016/j.camwa.2020.06.009}}

@article{venkat2025fft,
   title={Fast and scalable {FFT}-based {GPU}-accelerated algorithms for block-triangular {Toeplitz} matrices with application to linear inverse problems governed by autonomous dynamical systems},
   author={Sreeram Venkat and Milinda Fernando and Stefan Henneking and Omar Ghattas},
   journal = JSC,
   volume = {47},
   number = {5},
   pages = {B1201--B1226},
   year = {2025},
   doi = {10.1137/24M1683172}}

@article{henneking2026goal,
   title={Goal-oriented real-time {Bayesian} inference for linear autonomous dynamical systems with application to digital twins for tsunami early warning},
   author={Stefan Henneking and Sreeram Venkat and Omar Ghattas},
   journal=JCP,
   volume={552},
   pages={114682},
   year={2026},
   issn={0021-9991},
   doi={10.1016/j.jcp.2026.114682}}

@inproceedings{henneking2025bell,
   author = {Henneking, Stefan and Venkat, Sreeram and Dobrev, Veselin and Camier, John and Kolev, Tzanio and Fernando, Milinda and Gabriel, Alice-Agnes and Ghattas, Omar},
   title = {Real-Time {Bayesian} Inference at Extreme Scale: {A} Digital Twin for Tsunami Early Warning Applied to the {Cascadia Subduction Zone}},
   year = {2025},
   isbn = {9798400714665},
   publisher = {Association for Computing Machinery},
   address = {New York, NY, USA},
   doi = {10.1145/3712285.3771787},
   booktitle = {{Proceedings of the International Conference for High Performance Computing, Networking, Storage and Analysis}},
   pages = {60–71},
   numpages = {12},
   series = {SC '25}}

@Article{ghattas2021learning,
   title={Learning physics-based models from data: {P}erspectives from inverse problems and model reduction},
   author={Ghattas, Omar and Willcox, Karen},
   journal=ACTANUM,
   volume={30},
   pages={445--554},
   year={2021},
   publisher={Cambridge University Press},
   doi={10.1017/S0962492921000064}}

@Article{stuart2010inverse,
   title={{Inverse problems: A Bayesian perspective}},
   author={Stuart, Andrew M.},
   journal=ACTANUM,
   volume={19},
   pages={451--559},
   year={2010},
   publisher={Cambridge University Press},
   doi={10.1017/S0962492910000061}}

@article{alexanderian2016oed,
  title={On {Bayesian} {A}-and {D}-optimal experimental designs in infinite dimensions},
  author={Alexanderian, Alen and Gloor, Philip J and Ghattas, Omar},
  journal={Bayesian Analysis},
  volume={11},
  number={3},
  pages={671--695},
  year={2016},
  doi={10.1214/15-BA969}}

@book{atkinson2007oed,
   title={{Optimum Experimental Designs, with SAS}},
   author={Atkinson, Anthony and Donev, Alexander and Tobias, Randall},
   volume={34},
   year={2007},
   publisher={OUP Oxford},
   doi={10.1093/oso/9780199296590.001.0001}}

@book{pukelsheim2006oed,
   title={{Optimal Design of Experiments}},
   author={Pukelsheim, Friedrich},
   year={2006},
   publisher={SIAM},
   doi={10.1137/1.9780898719109}}

@article{attia2018goal,
   title={Goal-oriented optimal design of experiments for large-scale {Bayesian} linear inverse problems},
   author={Attia, Ahmed and Alexanderian, Alen and Saibaba, Arvind K.},
   journal={Inverse Problems},
   volume={34},
   number={9},
   pages={095009},
   year={2018},
   publisher={IOP Publishing},
   doi={10.1088/1361-6420/aad210}}

@article{wu2023goal,
   author = {Wu, Keyi and Chen, Peng and Ghattas, Omar},
   title = {An offline-online decomposition method for efficient linear {Bayesian} goal-oriented optimal experimental design: {Application} to optimal sensor placement},
   journal = JSC,
   volume = {45},
   number = {1},
   pages = {B57-B77},
   year = {2023},
   doi = {10.1137/21M1466542}}

@Book{KaipioSomersalo05,
    Title                    = {{Statistical and Computational Inverse Problems}},
    Author                   = {Kaipio, Jari and Somersalo, Erkki},
    Publisher                = {Springer-Verlag},
    Address                  = {New York, NY},
    Year                     = {2005},
    Series                   = {Applied Mathematical Sciences},
    Volume                   = {160},
    Doi                      = {10.1007/b138659}}

@article{Sun_2026_BackPropagatingEarthquakesSimple,
  title = {Back-propagating earthquakes on simple faults},
  author = {Sun, Yudong and Cattania, Camilla},
  year = 2026,
  journal = {AGU Advances},
  volume = {7},
  number = {1},
  pages = {e2025AV001649},
  issn = {2576-604X},
  doi = {10.1029/2025AV001649}
}

@article{Gabriel_2012_TransitionDynamicRupture,
  title = {The Transition of Dynamic Rupture Styles in Elastic Media under Velocity-Weakening Friction},
  author = {Gabriel, A.-A. and Ampuero, J.-P. and Dalguer, L. A. and Mai, P. M.},
  year = 2012,
  journal = {Journal of Geophysical Research: Solid Earth},
  volume = {117},
  number = {B9},
  issn = {2156-2202},
  doi = {10.1029/2012JB009468}
}

@article{Wirth_2025_ThreedimensionalSeismicVelocity,
  title = {Three-dimensional seismic velocity model for the {{Cascadia Subduction Zone}} with shallow soils and topography, version 1.7},
  author = {Wirth, Erin A. and Grant, Alex R. and Stone, Ian P. and Stephenson, William J. and Frankel, Arthur D.},
  year = 2025,
  journal = {U.S. Geological Survey Open-File Report 2025--1045},
  pages = {18},
  issn = {2331-1258},
  doi = {10.3133/ofr20251045}
}

@article{Tsuboi_1995_RapidDeterminationMw,
  title = {Rapid determination of {Mw} from broadband {P} waveforms},
  author = {Tsuboi, S. and Abe, K. and Takano, K. and Yamanaka, Y.},
  year = 1995,
  journal = {Bulletin of the Seismological Society of America},
  volume = {85},
  number = {2},
  pages = {606--613},
  issn = {0037-1106},
  doi = {10.1785/BSSA0850020606}
}

@article{Yao_2020_RuptureDynamics2012,
  title = {Rupture dynamics of the 2012 {Nicoya Mw} 7.6 Earthquake: {Evidence} for Low Strength on the Megathrust},
  author = {Yao, Suli and Yang, Hongfeng},
  year = 2020,
  journal = {Geophysical Research Letters},
  volume = {47},
  number = {13},
  pages = {e2020GL087508},
  issn = {1944-8007},
  doi = {10.1029/2020GL087508}
}

@article{Wang_2012_RealtimeForecastingApril,
  title = {Real-Time Forecasting of the {April} 11, 2012 {Sumatra} Tsunami},
  author = {Wang, Dailin and Becker, Nathan C. and Walsh, David and Fryer, Gerard J. and Weinstein, Stuart A. and McCreery, Charles S. and Sardi{\~n}a, Victor and Hsu, Vindell and Hirshorn, Barry F. and Hayes, Gavin P. and Duputel, Zacharie and Rivera, Luis and Kanamori, Hiroo and Koyanagi, Kanoa K. and Shiro, Brian},
  year = 2012,
  journal = {Geophysical Research Letters},
  volume = {39},
  number = {19},
  issn = {1944-8007},
  doi = {10.1029/2012GL053081}
}

@incollection{Talley_2011_ChapterGravityWaves,
  title = {Chapter 8 - {Gravity} Waves, Tides, and Coastal Oceanography},
  booktitle = {Descriptive {Physical Oceanography} ({Sixth Edition})},
  author = {Talley, Lynne D. and Pickard, George L. and Emery, William J. and Swift, James H.},
  editor = {Talley, Lynne D. and Pickard, George L. and Emery, William J. and Swift, James H.},
  year = 2011,
  pages = {223--244},
  publisher = {Academic Press},
  doi = {10.1016/B978-0-7506-4552-2.10008-3},
  isbn = {978-0-7506-4552-2}
}

@incollection{Kundu_2016_ChapterGravityWaves,
  title = {Chapter 8 - {Gravity Waves}},
  booktitle = {Fluid {Mechanics} ({Sixth Edition})},
  author = {Kundu, Pijush K. and Cohen, Ira M. and Dowling, David R.},
  editor = {Kundu, Pijush K. and Cohen, Ira M. and Dowling, David R.},
  year = 2016,
  pages = {349--407},
  publisher = {Academic Press},
  doi = {10.1016/B978-0-12-405935-1.00008-3},
  isbn = {978-0-12-405935-1}
}

@book{Levin_2015_PhysicsTsunamis,
  title = {Physics of {Tsunamis}},
  author = {Levin, Boris W. and Nosov, Mikhail A.},
  year = 2015,
  edition = {2},
  publisher = {Springer International Publishing},
  doi = {10.1007/978-3-319-24037-4},
  isbn = {978-3-319-24037-4}
}

@article{Tonegawa_2024_HighFrequencyTsunamisExcited,
  title = {High-frequency tsunamis excited near {Torishima Island}, {Japan}, observed by distributed acoustic sensing},
  author = {Tonegawa, T. and Araki, E.},
  year = 2024,
  journal = {Geophysical Research Letters},
  volume = {51},
  number = {11},
  pages = {e2024GL108714},
  issn = {1944-8007},
  doi = {10.1029/2024GL108714}
}

@article{Xiao_2024_DetectionEarthquakeInfragravity,
  title = {Detection of Earthquake Infragravity and Tsunami Waves With Underwater Distributed Acoustic Sensing},
  author = {Xiao, Han and Spica, Zack J. and Li, Jiaxuan and Zhan, Zhongwen},
  year = 2024,
  journal = {Geophysical Research Letters},
  volume = {51},
  number = {2},
  pages = {e2023GL106767},
  issn = {1944-8007},
  doi = {10.1029/2023GL106767}
}

@article{Witter_2013_SimulatedTsunamiInundation,
  title = {Simulated Tsunami Inundation for a Range of {Cascadia} Megathrust Earthquake Scenarios at {Bandon}, {Oregon}, {USA}},
  author = {Witter, Robert C. and Zhang, Yinglong J. and Wang, Kelin and Priest, George R. and Goldfinger, Chris and Stimely, Laura and English, John T. and Ferro, Paul A.},
  year = 2013,
  journal = {Geosphere},
  volume = {9},
  number = {6},
  pages = {1783--1803},
  issn = {1553-040X},
  doi = {10.1130/GES00899.1}
}

@article{Savage_1964_PropertiesTensileFractures,
  title = {Some {Properties} of {Tensile Fractures Inferred} from {Elastic Wave Radiation}},
  author = {Savage, J. C. and Hasegawa, H. S.},
  year = 1964,
  journal = {Journal of Geophysical Research (1896-1977)},
  volume = {69},
  number = {10},
  pages = {2091--2100},
  issn = {2156-2202},
  doi = {10.1029/JZ069i010p02091}
}

@article{Xu_2023_2016MenyuanEarthquake,
  title = {The 2016 {Menyuan} Earthquake: {The} Largest Self-Arrested Crustal Earthquake Ever Observed},
  author = {Xu, Duyuan and Gong, Wenzheng and Zhang, Zhenguo and Xu, Jiankuan and Yu, Houyun and Chen, Xiaofei},
  year = 2023,
  journal = {Geophysical Research Letters},
  volume = {50},
  number = {11},
  pages = {e2023GL103556},
  issn = {1944-8007},
  doi = {10.1029/2023GL103556}
}

@article{Kame_2003_DynamicBranchingArresting,
  title = {Dynamic Branching, Arresting of Rupture and the Seismic Wave Radiation in Self-Chosen Crack Path Modelling},
  author = {Kame, Nobuki and Yamashita, Teruo},
  year = 2003,
  journal = {Geophysical Journal International},
  volume = {155},
  number = {3},
  pages = {1042--1050},
  issn = {0956-540X},
  doi = {10.1111/j.1365-246X.2003.02113.x}
}

@article{Mosconi_2026_DiscriminatingDynamicRupture,
  title = {Discriminating Dynamic Rupture Arrest in Fluid-Induced Microearthquakes Using Spectral Inversion},
  author = {Mosconi, Francesco and Tinti, Elisa and Supino, Mariano and Gabriel, Alice-Agnes and Casarotti, Emanuele and Meier, Men-Andrin and Rinaldi, Antonio Pio and Giardini, Domenico and Cocco, Massimo},
  year = 2026,
  journal = {Earth and Planetary Science Letters},
  volume = {678},
  pages = {119816},
  issn = {0012-821X},
  doi = {10.1016/j.epsl.2025.119816}
}

@article{Ding_2024_BackPropagatingRuptureNature,
  title = {Back-propagating rupture: {Nature}, excitation, and implications},
  author = {Ding, Xiaotian and Xu, Shiqing and Fukuyama, Eiichi and Yamashita, Futoshi},
  year = 2024,
  journal = {Journal of Geophysical Research: Solid Earth},
  volume = {129},
  number = {10},
  pages = {e2024JB029629},
  issn = {2169-9356},
  doi = {10.1029/2024JB029629}
}

@article{Kearse_2026_StoppingPhaseReveals,
  title = {Stopping Phase Reveals Abrupt Arrest of Large Strike-Slip Earthquakes},
  author = {Kearse, Jesse and Kaneko, Yoshihiro},
  year = 2026,
  journal = {Science},
  volume = {0},
  number = {0},
  pages = {eaef3733},
  publisher = {American Association for the Advancement of Science},
  doi = {10.1126/science.aef3733}
}

@article{Madariaga_1976_DynamicsExpandingCircular,
  title = {Dynamics of an expanding circular fault},
  author = {Madariaga, Raul},
  year = 1976,
  journal = {Bulletin of the Seismological Society of America},
  volume = {66},
  number = {3},
  pages = {639--666},
  issn = {0037-1106},
  doi = {10.1785/BSSA0660030639}
}

@article{Schliwa_2023_EquivalentNearFieldCorner,
  title = {Equivalent near-field corner frequency analysis of 3D dynamic rupture simulations reveals dynamic source effects},
  author = {Schliwa, Nico and Gabriel, Alice-Agnes},
  year = 2023,
  journal = {Seismological Research Letters},
  volume = {95},
  number = {2A},
  pages = {900--924},
  issn = {0895-0695},
  doi = {10.1785/0220230225}
}

@article{Savage_1965_StoppingPhaseSeismograms,
  title = {The stopping phase on seismograms},
  author = {Savage, J. C.},
  year = 1965,
  journal = {Bulletin of the Seismological Society of America},
  volume = {55},
  number = {1},
  pages = {47--58},
  issn = {0037-1106},
  doi = {10.1785/BSSA0550010047}
}

@article{Press_1950_AiryPhaseShallowfocus,
  title = {The airy phase of shallow-focus submarine earthquakes},
  author = {Press, Frank and Ewing, Maurice and Tolstoy, Ivan},
  year = 1950,
  journal = {Bulletin of the Seismological Society of America},
  volume = {40},
  number = {2},
  pages = {111--148},
  issn = {0037-1106},
  doi = {10.1785/BSSA0400020111}
}

@incollection{Pekeris_1948_TheoryPropagationExplosive,
  title = {Theory of propagation of explosive sounds in shallow water},
  booktitle = {Propagation of {{Sound}} in the {{Ocean}}},
  author = {Pekeris, C. L.},
  editor = {Worzel, J. Lamar and Ewing, Maurice and Pekeris, C. L.},
  year = 1948,
  volume = {27},
  publisher = {Geological Society of America},
  doi = {10.1130/MEM27-2-p1},
  isbn = {978-0-8137-1027-3}
}

@article{Wirp_2021_3DLinkedSubduction,
  title = {{3D} linked subduction, dynamic rupture, tsunami, and inundation modeling: {Dynamic} effects of supershear and tsunami earthquakes, hypocenter location, and shallow fault slip},
  author = {Wirp, Sara Aniko and Gabriel, Alice Agnes and Schmeller, Maximilian and H. Madden, Elizabeth and {van Zelst}, Iris and Krenz, Lukas and {van Dinther}, Ylona and Rannabauer, Leonhard},
  year = 2021,
  journal = {Frontiers in Earth Science},
  volume = {9},
  pages = {177},
  publisher = {Frontiers Media S.A.},
  issn = {22966463},
  doi = {10.3389/feart.2021.626844}
}

@article{Becerril_2026_TsunamiEarlyWarningDistributed,
  title = {Towards tsunami early-warning with distributed acoustic sensing: {Expected} seafloor strains induced by tsunamis},
  author = {Becerril, Carlos and Sladen, Anthony and Ampuero, Jean-Paul and {Preciado-Garbayo}, Javier and {Gonzalez-Herraez}, Miguel and Kutschera, Fabian and Gabriel, Alice-Agnes and Bouchette, Frederic},
  year = 2026,
  journal = {Pure and Applied Geophysics},
  issn = {1420-9136},
  doi = {10.1007/s00024-026-03940-1}
}

@article{Nakanishi_1992_RayleighWavesGuided,
  title = {Rayleigh Waves Guided by Sea-Trench Topography},
  author = {Nakanishi, Ichiro},
  year = 1992,
  journal = {Geophysical Research Letters},
  volume = {19},
  number = {24},
  pages = {2385--2388},
  issn = {1944-8007},
  doi = {10.1029/92GL02438}
}

@article{Biot_1952_InteractionRayleighStoneley,
  title = {The Interaction of {{Rayleigh}} and {{Stoneley}} Waves in the Ocean Bottom},
  author = {Biot, M. A.},
  year = 1952,
  journal = {Bulletin of the Seismological Society of America},
  volume = {42},
  number = {1},
  pages = {81--93},
  issn = {0037-1106},
  doi = {10.1785/BSSA0420010081}
}

@article{Saito_2019_SynthesizingSeaSurface,
  title = {Synthesizing Sea Surface Height Change Including Seismic Waves and Tsunami Using a Dynamic Rupture Scenario of Anticipated {{Nankai}} Trough Earthquakes},
  author = {Saito, Tatsuhiko and Baba, Toshitaka and Inazu, Daisuke and Takemura, Shunsuke and Fukuyama, Eiichi},
  year = 2019,
  journal = {Tectonophysics},
  volume = {769},
  pages = {228166},
  publisher = {Elsevier},
  issn = {0040-1951},
  doi = {10.1016/j.tecto.2019.228166}
}

@incollection{Hirshorn_2020_EarthquakeSourceParameters,
  title = {Earthquake source parameters, rapid estimates for tsunami forecasts and warnings},
  booktitle = {Encyclopedia of {{Complexity}} and {{Systems Science}}},
  author = {Hirshorn, Barry and Weinstein, Stuart and Wang, Dailin and Koyanagi, Kanoa and Becker, Nathan and McCreery, Charles},
  editor = {Meyers, Robert A.},
  year = 2020,
  pages = {1--35},
  publisher = {Springer},
  doi = {10.1007/978-3-642-27737-5_160-2},
  isbn = {978-3-642-27737-5}
}

@misc{Henneking_2026_SupplementaryMaterialRealtime,
  title = {Supplementary Material for "{Real-time} Probabilistic Tsunami Forecasting in {Cascadia} from Sparse Offshore Pressure Observations"},
  author = {Henneking, Stefan and Kutschera, Fabian and Venkat, Sreeram and Gabriel, Alice-Agnes and Ghattas, Omar},
  year = 2026,
  publisher = {Zenodo},
  doi = {10.5281/zenodo.19141625}
}

@article{Abdolali_2015_HydroacousticTsunamiWaves,
  title = {Hydro-acoustic and tsunami waves generated by the 2012 {{Haida Gwaii}} earthquake: {{Modeling}} and in situ measurements},
  author = {Abdolali, Ali and Cecioni, Claudia and Bellotti, Giorgio and Kirby, James T.},
  year = 2015,
  journal = {Journal of Geophysical Research: Oceans},
  volume = {120},
  number = {2},
  pages = {958--971},
  issn = {2169-9291},
  doi = {10.1002/2014JC010385}
}

@article{Abrahams_2023_ComparisonMethodsCoupled,
  title = {Comparison of Methods for Coupled Earthquake and Tsunami Modelling},
  author = {Abrahams, Lauren S and Krenz, Lukas and Dunham, Eric M and Gabriel, Alice-Agnes and Saito, Tatsuhiko},
  year = 2023,
  journal = {Geophysical Journal International},
  volume = {234},
  number = {1},
  pages = {404--426},
  issn = {0956-540X},
  doi = {10.1093/gji/ggad053}
}

@article{Andrews_1976_RuptureVelocity_LSW,
  title = {Rupture Velocity of Plane Strain Shear Cracks},
  author = {Andrews, D. J.},
  year = 1976,
  journal = {Journal of Geophysical Research},
  volume = {81},
  number = {32},
  pages = {5679--5687},
  publisher = {John Wiley \& Sons, Ltd},
  issn = {2156-2202},
  doi = {10.1029/JB081I032P05679}
}

@article{Aoi_2020_MOWLASNIEDObservation,
  title = {{{MOWLAS}}: {{NIED}} Observation Network for Earthquake, Tsunami and Volcano},
  author = {Aoi, Shin and Asano, Youichi and Kunugi, Takashi and Kimura, Takeshi and Uehira, Kenji and Takahashi, Narumi and Ueda, Hideki and Shiomi, Katsuhiko and Matsumoto, Takumi and Fujiwara, Hiroyuki},
  year = 2020,
  journal = {Earth, Planets and Space},
  volume = {72},
  number = {1},
  pages = {126},
  issn = {1880-5981},
  doi = {10.1186/s40623-020-01250-x}
}

@article{Atwater_1991_SuddenProbablyCoseismic,
  title = {Sudden, Probably Coseismic Submergence of {{Holocene}} Trees and Grass in Coastal {{Washington State}}},
  author = {Atwater, Brian F. and Yamaguchi, David K.},
  year = 1991,
  journal = {Geology},
  volume = {19},
  number = {7},
  pages = {706--709},
  issn = {0091-7613},
  doi = {10.1130/0091-7613(1991)019<0706:SPCSOH>2.3.CO;2}
}

@article{Atwater_1995_SummaryCoastalGeologic,
  title = {Summary of Coastal Geologic Evidence for Past Great Earthquakes at the {Cascadia Subduction Zone}},
  author = {Atwater, Brian F. and Nelson, Alan R. and Clague, John J. and Carver, Gary A. and Yamaguchi, David K. and Bobrowsky, Peter T. and Bourgeois, Joanne and Darienzo, Mark E. and Grant, Wendy C. and {Hemphill-Haley}, Eileen and Kelsey, Harvey M. and Jacoby, Gordon C. and Nishenko, Stuart P. and Palmer, Stephen P. and Peterson, Curt D. and Reinhart, Mary Ann},
  year = 1995,
  journal = {Earthquake Spectra},
  volume = {11},
  number = {1},
  pages = {1--18},
  publisher = {SAGE Publications Ltd STM},
  issn = {8755-2930},
  doi = {10.1193/1.1585800}
}

@article{Beresnev_2003_UncertaintiesFiniteFaultSlip,
  title = {Uncertainties in Finite-Fault Slip Inversions: {To} What Extent to Believe? (A Critical Review)},
  author = {Beresnev, Igor A.},
  year = 2003,
  journal = {Bulletin of the Seismological Society of America},
  volume = {93},
  number = {6},
  pages = {2445--2458},
  issn = {0037-1106},
  doi = {10.1785/0120020225}
}

@article{Beyreuther_2010_ObsPy,
  title = {{{ObsPy}}: {A} {Python} Toolbox for Seismology},
  author = {Beyreuther, Moritz and Barsch, Robert and Krischer, Lion and Megies, Tobias and Behr, Yannik and Wassermann, Joachim},
  year = 2010,
  journal = {Seismological Research Letters},
  volume = {81},
  number = {3},
  pages = {530--533},
  publisher = {GeoScienceWorld},
  issn = {0895-0695},
  doi = {10.1785/gssrl.81.3.530}
}

@article{Biemiller_2025_StructuralControlsSplay,
  title = {Structural Controls on Splay Fault Rupture Dynamics During {Cascadia} Megathrust Earthquakes},
  author = {Biemiller, J. and Gabriel, A.-A. and Staisch, L. and Ulrich, T. and Dunham, A. and Wirth, E. and Watt, J. and Lucas, M. C. and Ledeczi, A.},
  year = 2025,
  journal = {AGU Advances},
  volume = {6},
  number = {6},
  pages = {e2025AV001812},
  issn = {2576-604X},
  doi = {10.1029/2025AV001812}
}

@article{Bird_2003_UpdatedDigitalModel,
  title = {An Updated Digital Model of Plate Boundaries},
  author = {Bird, Peter},
  year = 2003,
  journal = {Geochemistry, Geophysics, Geosystems},
  volume = {4},
  number = {3},
  pages = {2001GC000252},
  issn = {1525-2027, 1525-2027},
  doi = {10.1029/2001GC000252}
}

@article{Carbotte_2024_SubductingPlateStructure,
  title = {Subducting Plate Structure and Megathrust Morphology from Deep Seismic Imaging Linked to Earthquake Rupture Segmentation at {{Cascadia}}},
  author = {Carbotte, Suzanne M. and Boston, Brian and Han, Shuoshuo and Shuck, Brandon and Beeson, Jeffrey and Canales, J. Pablo and Tobin, Harold and Miller, Nathan and Nedimovic, Mladen and Tr{\'e}hu, Anne and Lee, Michelle and Lucas, Madelaine and Jian, Hanchao and Jiang, Danqi and Moser, Liam and Anderson, Chris and Judd, Darren and Fernandez, Jaime and Campbell, Chuck and Goswami, Antara and Gahlawat, Rajendra},
  year = 2024,
  journal = {Science Advances},
  volume = {10},
  number = {23},
  pages = {eadl3198},
  doi = {10.1126/sciadv.adl3198}
}

@article{Crotwell_1999_TauPToolkitFlexible,
  title = {The {{TauP toolkit}}: {Flexible} Seismic Travel-time and Ray-path Utilities},
  author = {Crotwell, H. Philip and Owens, Thomas J. and Ritsema, Jeroen},
  year = 1999,
  journal = {Seismological Research Letters},
  volume = {70},
  number = {2},
  pages = {154--160},
  issn = {0895-0695},
  doi = {10.1785/gssrl.70.2.154}
}

@article{Crowell_2016_DemonstrationCascadiaGFAST,
  title = {Demonstration of the {{Cascadia G}}-{{FAST}} Geodetic Earthquake Early Warning System for the {{Nisqually}}, {{Washington}}, {{Earthquake}}},
  author = {Crowell, Brendan W. and Schmidt, David A. and Bodin, Paul and Vidale, John E. and Gomberg, Joan and Renate Hartog, J. and Kress, Victor C. and Melbourne, Timothy I. and Santillan, Marcelo and Minson, Sarah E. and Jamison, Dylan G.},
  year = 2016,
  journal = {Seismological Research Letters},
  volume = {87},
  number = {4},
  pages = {930--943},
  issn = {0895-0695},
  doi = {10.1785/0220150255}
}

@article{Day_1982_ThreedimensionalSimulationSpontaneous,
  title = {Three-dimensional simulation of spontaneous rupture: {{The}} effect of nonuniform prestress},
  author = {Day, Steven M.},
  year = 1982,
  journal = {Bulletin of the Seismological Society of America},
  volume = {72},
  number = {6A},
  pages = {1881--1902},
  issn = {0037-1106},
  doi = {10.1785/BSSA07206A1881}
}

@article{DeSanto_2025_FullLockingShallow,
  title = {Near Full Locking on the Shallow Megathrust of the Central {Cascadia Subduction Zone} Revealed by {{GNSS-Acoustic}}},
  author = {DeSanto, John B. and Schmidt, David A. and Zumberge, Mark and Sasagawa, Glenn and Chadwell, C. David},
  year = 2025,
  journal = {Earth and Planetary Science Letters},
  volume = {665},
  pages = {119463},
  issn = {0012-821X},
  doi = {10.1016/j.epsl.2025.119463}
}

@article{Du_2025_WedgeInelasticityFully,
  title = {Wedge inelasticity and fully coupled models of dynamic rupture, ocean acoustic waves, and tsunami in the {Japan Trench}: {The} 1896 {Sanriku} Earthquake},
  author = {Du, Yue and Ma, Shuo},
  year = 2025,
  journal = {Journal of Geophysical Research: Solid Earth},
  volume = {130},
  number = {10},
  pages = {e2025JB032410},
  issn = {2169-9356},
  doi = {10.1029/2025JB032410}
}

@article{Dumbser_2006_ArbitraryHighorderDiscontinuous,
  title = {An Arbitrary High-Order Discontinuous {{Galerkin}} Method for Elastic Waves on Unstructured Meshes - {{II}}. {{The}} Three-Dimensional Isotropic Case},
  author = {Dumbser, M and K{\"a}ser, M},
  year = 2006,
  journal = {Geophysical Journal International},
  volume = {167},
  number = {1},
  pages = {319--336},
  doi = {10.1111/j.1365-246X.2006.03120.x}
}

@article{Dunham_2025_Impact3DStructure,
  title = {The impact of {3D} structure on coseismic coastal land-level change and tsunami generation in the {Cascadia Subduction Zone}},
  author = {Dunham, Audrey M. and Kim, Jeonghyeop and Wirth, Erin and Schmidt, David and LeVeque, Randall J. and Wei, Yong and Adams, Loyce M. and Pollitz, Fred},
  year = 2025,
  journal = {Geophysical Research Letters},
  volume = {52},
  number = {24},
  pages = {e2025GL117783},
  issn = {1944-8007},
  doi = {10.1029/2025GL117783}
}

@article{Duputel_2014_AccountingPredictionUncertainty,
  title = {Accounting for Prediction Uncertainty When Inferring Subsurface Fault Slip},
  author = {Duputel, Zacharie and Agram, Piyush S. and Simons, Mark and Minson, Sarah E. and Beck, James L.},
  year = 2014,
  journal = {Geophysical Journal International},
  volume = {197},
  number = {1},
  pages = {464--482},
  issn = {0956-540X},
  doi = {10.1093/gji/ggt517}
}

@article{Dushaw_1993_EquationsSpeedSound,
  title = {On Equations for the Speed of Sound in Seawater},
  author = {Dushaw, Brian D. and Worcester, Peter F. and Cornuelle, Bruce D. and Howe, Bruce M.},
  year = 1993,
  journal = {The Journal of the Acoustical Society of America},
  volume = {93},
  number = {1},
  pages = {255--275},
  issn = {0001-4966},
  doi = {10.1121/1.405660}
}

@misc{Gabriel_2025_SeisSol,
  title = {{{SeisSol}}},
  author = {Gabriel, Alice-Agnes and Kurapati, Vikas and Niu, Zihua and Schliwa, Nico and Schneller, David and Ulrich, Thomas and Dorozhinskii, Ravil and Krenz, Lukas and Uphoff, Carsten and Wolf, Sebastian and Breuer, Alexander and Heinecke, Alexander and Pelties, Christian and Rettenberger, Sebastian and Wollherr, Stephanie and Bader, Michael},
  year = 2025,
  doi = {10.5281/zenodo.4672483},
  howpublished = {Zenodo}
}

@misc{Gebco_2020,
  title = {{{GEBCO}} 2020 {{Grid}}},
  author = {{GEBCO Bathymetric Compilation Group}},
  year = 2020,
  doi = {10.5285/a29c5465-b138-234d-e053-6c86abc040b9}
}

@article{Glehman_2025_PartialRupturesGoverned,
  title = {Partial Ruptures Governed by the Complex Interplay between Geodetic Slip Deficit, Rigidity, and Pore Fluid Pressure in {{3D Cascadia}} Dynamic Rupture Simulations},
  author = {Glehman, Jonatan and Gabriel, Alice and Ulrich, Thomas and Ramos, Marlon and Huang, Yihe and Lindsey, Eric},
  year = 2025,
  journal = {Seismica},
  volume = {2},
  number = {4},
  issn = {2816-9387},
  doi = {10.26443/seismica.v2i4.1427}
}

@article{Goldfinger_2025_UnravellingDanceEarthquakes,
  title = {Unravelling the Dance of Earthquakes: {{Evidence}} of Partial Synchronization of the Northern {{San Andreas}} Fault and {{Cascadia}} Megathrust},
  author = {Goldfinger, C. and Beeson, J. and Black, B. and Vizcaino, A. and Nelson, C.H. and Morey, A. and Patton, J.R. and {Guti{\'e}rrez-Pastor}, J. and Romsos, C. and Walzcak, M.D.},
  year = 2025,
  journal = {Geosphere},
  issn = {1553-040X},
  doi = {10.1130/GES02857.1}
}

@article{Golriz_2023_RealTimeSeismogeodeticEarthquake,
  title = {Real-time seismogeodetic earthquake magnitude estimates for local tsunami warnings},
  author = {Golriz, Dorian and Hirshorn, Barry and Bock, Yehuda and Weinstein, Stuart and Weiss, Jonathan R.},
  year = 2023,
  journal = {Journal of Geophysical Research: Solid Earth},
  volume = {128},
  number = {1},
  pages = {e2022JB025555},
  issn = {2169-9356},
  doi = {10.1029/2022JB025555}
}

@inproceedings{Gonzalez_1998_DeepoceanAssessmentReporting,
  title = {Deep-ocean assessment and reporting of tsunamis ({DART}): {Brief} overview and status report},
  booktitle = {Proceedings of the International Workshop on Tsunami Disaster Mitigation},
  author = {Gonzalez, Frank I and Milburn, Hank M and Bernard, Eddie N and Newman, Jean C},
  year = 1998,
  publisher = {Tokyo, Japan}
}

@article{Gusman_2014_PhaseInversionTsunami,
  title = {W phase inversion and tsunami inundation modeling for tsunami early warning: {Case} study for the 2011 {Tohoku} event},
  author = {Gusman, Aditya Riadi and Tanioka, Yuichiro},
  year = 2014,
  journal = {Pure and Applied Geophysics},
  volume = {171},
  number = {7},
  pages = {1409--1422},
  issn = {1420-9136},
  doi = {10.1007/s00024-013-0680-z}
}

@article{Harris_2018_SuiteExercisesVerifying,
  title = {A Suite of Exercises for Verifying Dynamic Earthquake Rupture Codes},
  author = {Harris, Ruth A. and Aagaard, Brad and Barall, Michael and Ma, Shuo and Roten, Daniel and Olsen, Kim and Duan, Benchun and Liu, Dunyu and Luo, Bin and Bai, Kangchen and Ampuero, Jean Paul and Kaneko, Yoshihiro and Gabriel, Alice Agnes and Duru, Kenneth and Ulrich, Thomas and Wollherr, Stephanie and Shi, Zheqiang and Dunham, Eric and Bydlon, Sam and Zhang, Zhenguo and Chen, Xiaofei and Somala, Surendra Nadh and Pelties, Christian and Tago, Josu{\'e} and {Cruz-Atienza}, Victor Manuel and Kozdon, Jeremy and Daub, Eric and Aslam, Khurram and Kase, Yuko and Withers, Kyle and Dalguer, Luis},
  year = 2018,
  journal = {Seismological Research Letters},
  volume = {89},
  number = {3},
  pages = {1146--1162},
  issn = {1938-2057},
  doi = {10.1785/0220170222}
}

@article{Hayes_2018_Slab2ComprehensiveSubduction,
  title = {Slab2, a Comprehensive Subduction Zone Geometry Model},
  author = {Hayes, Gavin P. and Moore, Ginevra L. and Portner, Daniel E. and Hearne, Mike and Flamme, Hanna and Furtney, Maria and Smoczyk, Gregory M.},
  year = 2018,
  journal = {Science},
  volume = {362},
  number = {6410},
  pages = {58--61},
  publisher = {American Association for the Advancement of Science},
  doi = {10.1126/science.aat4723}
}

@article{Ida_1972_CohesiveForce_LSW,
  title = {Cohesive Force across the Tip of a Longitudinal-Shear Crack and {{Griffith}}'s Specific Surface Energy},
  author = {Ida, Yoshiaki},
  year = 1972,
  journal = {Journal of Geophysical Research},
  volume = {77},
  number = {20},
  pages = {3796--3805},
  publisher = {John Wiley \& Sons, Ltd},
  issn = {2156-2202},
  doi = {10.1029/JB077I020P03796}
}

@article{Kanamori_2008_SourceInversionOfWphasea,
  title = {Source inversion of {W} phase: {Speeding} up seismic tsunami warning},
  author = {Kanamori, Hiroo and Rivera, Luis},
  year = 2008,
  journal = {Geophysical Journal International},
  volume = {175},
  number = {1},
  pages = {222--238},
  issn = {0956-540X},
  doi = {10.1111/j.1365-246X.2008.03887.x}
}

@inproceedings{Kanazawa_2013_JapanTrenchEarthquake,
  title = {Japan {{Trench}} Earthquake and Tsunami Monitoring Network of Cable-Linked 150 Ocean Bottom Observatories and Its Impact to Earth Disaster Science},
  booktitle = {2013 {{IEEE International Underwater Technology Symposium}} ({{UT}})},
  author = {Kanazawa, Toshihiko},
  year = 2013,
  pages = {1--5},
  doi = {10.1109/UT.2013.6519911}
}

@article{Kaser_2006_ArbitraryHighorderDiscontinuous,
  title = {An Arbitrary High-Order Discontinuous {{Galerkin}} Method for Elastic Waves on Unstructured Meshes - {{I}}. {{The}} Two-Dimensional Isotropic Case with External Source Terms},
  author = {K{\"a}ser, Martin and Dumbser, Michael},
  year = 2006,
  journal = {Geophysical Journal International},
  volume = {166},
  number = {2},
  pages = {855--877},
  issn = {0956540X},
  doi = {10.1111/j.1365-246X.2006.03051.x}
}

@article{Kemp_2018_RevisingEstimatesSpatially,
  title = {Revising estimates of spatially variable subsidence during the {{A}}.{{D}}. 1700 {Cascadia} earthquake using a {Bayesian} foraminiferal transfer function},
  author = {Kemp, Andrew C. and Cahill, Niamh and Engelhart, Simon E. and Hawkes, Andrea D. and Wang, Kelin},
  year = 2018,
  journal = {Bulletin of the Seismological Society of America},
  volume = {108},
  number = {2},
  pages = {654--673},
  issn = {0037-1106},
  doi = {10.1785/0120170269}
}

@article{Kozdon_2014_ConstrainingShallowSlip,
  title = {Constraining Shallow Slip and Tsunami Excitation in Megathrust Ruptures Using Seismic and Ocean Acoustic Waves Recorded on Ocean-Bottom Sensor Networks},
  author = {Kozdon, Jeremy E. and Dunham, Eric M.},
  year = 2014,
  journal = {Earth and Planetary Science Letters},
  volume = {396},
  pages = {56--65},
  issn = {0012-821X},
  doi = {10.1016/j.epsl.2014.04.001}
}

@Inproceedings{Krenz_2021_FullyCoupled,
   title={{3D} acoustic-elastic coupling with gravity: {The} dynamics of the 2018 {Palu, Sulawesi} earthquake and tsunami},
   author={Krenz, Lukas and Uphoff, Carsten and Ulrich, Thomas and Gabriel, Alice-Agnes and Abrahams, Lauren S. and Dunham, Eric M. and Bader, Michael},
   booktitle={Proceedings of the International Conference for High Performance Computing, Networking, Storage and Analysis},
   pages={1--14},
   year={2021},
   doi={10.1145/3458817.3476173}}

@inproceedings{Kutschera_2025_CRESCENTSCECUSGS,
  title = {The {{CRESCENT}}/{{SCEC}}/{{USGS}} Tsunami Benchmarks: {3D} Fully Coupled Earthquake Dynamic Rupture and Tsunami Simulations with Varying Bathymetric Complexity},
  booktitle = {Abstract ({{T43E-0182}}) Presented at {{AGU25}}, 15-19 {{Dec}}},
  author = {Kutschera, Fabian and Gabriel, Alice-Agnes and Dunham, Eric M and Harris, Ruth and Barall, Michael and Bachelot, Lo{\"i}c and Ma, Shuo and Schneller, David and Zhang, Wenqiang},
  year = 2025,
  publisher = {AGU Annual Meeting 2025}
}

@article{Kwong_2025_PerformanceSlabGeometry,
  title = {Performance of Slab Geometry Constraints on Rapid Geodetic Slip Models, Tsunami Amplitude, and Inundation Estimates in {{Cascadia}}},
  author = {Kwong, Kevin and Crowell, Brendan and Williamson, Amy and Melgar, Diego and Wei, Yong},
  year = 2025,
  journal = {Seismica},
  volume = {2},
  number = {4},
  issn = {2816-9387},
  doi = {10.26443/seismica.v2i4.1406}
}

@Article{Liu_2021_ComparisonMachineLearning,
    title={Comparison of machine learning approaches for tsunami forecasting from sparse observations},
    author={Liu, Christopher M. and Rim, Donsub and Baraldi, Robert and LeVeque, Randall J.},
    journal=PAGEO,
    volume={178},
    number={12},
    pages={5129--5153},
    year={2021},
    publisher={Springer},
    doi={10.1007/s00024-021-02841-9}}

@Article{Lotto_2015_HighorderFiniteDifference,
   title={High-order finite difference modeling of tsunami generation in a compressible ocean from offshore earthquakes},
   author={Lotto, Gabriel C. and Dunham, Eric M.},
   journal=CGEO,
   volume={19},
   number={2},
   pages={327--340},
   year={2015},
   publisher={Springer},
   doi={10.1007/s10596-015-9472-0}}

@article{Lotto_2018_AppliedFullyCoupled,
  title = {Fully coupled simulations of megathrust earthquakes and tsunamis in the {Japan Trench}, {Nankai Trough}, and {Cascadia Subduction Zone}},
  author = {Lotto, Gabriel C. and Jeppson, Tamara N. and Dunham, Eric M.},
  year={2019},
  journal = {Pure and Applied Geophysics},
  volume = {176},
  number = {9},
  pages = {4009--4041},
  doi = {10.1007/S00024-018-1990-Y}}

@article{Lucas_2023_NoEvidenceActive,
  title = {No Evidence for an Active Margin-Spanning Megasplay Fault at the {{Cascadia Subduction Zone}}},
  author = {Lucas, Madeleine C. and Ledeczi, Anna M. and Tobin, Harold J. and Carbotte, Suzanne M. and Watt, Janet T. and Han, Shuoshuo and Boston, Brian and Jiang, Danqi},
  year = 2023,
  journal = {Seismica},
  volume = {2},
  number = {4},
  issn = {2816-9387},
  doi = {10.26443/seismica.v2i4.1477}
}

@article{Ma_2025_WedgeInelasticityFully,
  title = {Wedge Inelasticity and Fully Coupled Models of Dynamic Rupture, Ocean Acoustic Waves, and Tsunami in the {{Japan Trench}}: {{The}} 2011 {{Tohoku-Oki}} Earthquake},
  author = {Ma, Shuo and Du, Yue},
  year = 2025,
  journal = {Tectonophysics},
  volume = {910},
  pages = {230831},
  issn = {0040-1951},
  doi = {10.1016/j.tecto.2025.230831}
}

@article{Mackenzie_1981_NinetermEquationSound,
  title = {Nine-term Equation for Sound Speed in the Oceans},
  author = {Mackenzie, Kenneth V.},
  year = 1981,
  journal = {The Journal of the Acoustical Society of America},
  volume = {70},
  number = {3},
  pages = {807--812},
  issn = {0001-4966},
  doi = {10.1121/1.386920}
}

@article{Madden_2022_StatePoreFluid,
  title = {The state of pore fluid pressure and 3-{D} megathrust earthquake dynamics},
  author = {Madden, Elizabeth H. and Ulrich, Thomas and Gabriel, Alice-Agnes},
  year = 2022,
  journal = {Journal of Geophysical Research: Solid Earth},
  volume = {127},
  number = {4},
  pages = {e2021JB023382},
  issn = {2169-9356},
  doi = {10.1029/2021JB023382}
}

@article{Maeda_2013_FDMSimulationSeismic,
  title = {{FDM} simulation of seismic waves, ocean acoustic waves, and tsunamis based on tsunami-coupled equations of motion},
  author = {Maeda, Takuto and Furumura, Takashi},
  year = 2013,
  journal = {Pure and Applied Geophysics},
  volume = {170},
  number = {1},
  pages = {109--127},
  issn = {1420-9136},
  doi = {10.1007/s00024-011-0430-z}
}

@article{Maeda_2013_SeismicTsunamiWave,
  title = {Seismic- and tsunami-wave propagation of the 2011 off the {Pacific Coast} of {Tohoku} earthquake as inferred from the tsunami-coupled finite-difference simulation},
  author = {Maeda, Takuto and Furumura, Takashi and Noguchi, Shinako and Takemura, Shunsuke and Sakai, Shin'ichi and Shinohara, Masanao and Iwai, Kazuhisa and Lee, Shiann Jong},
  year = 2013,
  journal = {Bulletin of the Seismological Society of America},
  volume = {103},
  number = {2B},
  pages = {1456--1472},
  publisher = {GeoScienceWorld},
  issn = {0037-1106},
  doi = {10.1785/0120120118}
}

@article{Melgar_2013_FieldTsunamiModels,
  title = {Near-Field Tsunami Models with Rapid Earthquake Source Inversions from Land- and Ocean-Based Observations: {{The}} Potential for Forecast and Warning},
  author = {Melgar, Diego and Bock, Yehuda},
  year = 2013,
  journal = {Journal of Geophysical Research: Solid Earth},
  volume = {118},
  number = {11},
  pages = {5939--5955},
  issn = {2169-9356},
  doi = {10.1002/2013JB010506}
}

@article{Melgar_2016_LocalTsunamiWarnings,
  title = {Local Tsunami Warnings: {{Perspectives}} from Recent Large Events},
  author = {Melgar, Diego and Allen, Richard M. and Riquelme, Sebastian and Geng, Jianghui and Bravo, Francisco and Baez, Juan Carlos and Parra, Hector and Barrientos, Sergio and Fang, Peng and Bock, Yehuda and Bevis, Michael and Caccamise II, Dana J. and Vigny, Christophe and Moreno, Marcos and Smalley Jr., Robert},
  year = 2016,
  journal = {Geophysical Research Letters},
  volume = {43},
  number = {3},
  pages = {1109--1117},
  issn = {1944-8007},
  doi = {10.1002/2015GL067100}
}

@article{Mizutani_2020_EarlyTsunamiDetection,
  title = {Early tsunami detection with near-fault ocean-bottom pressure gauge records based on the comparison with seismic data},
  author = {Mizutani, Ayumu and Yomogida, Kiyoshi and Tanioka, Yuichiro},
  year = 2020,
  journal = {Journal of Geophysical Research: Oceans},
  volume = {125},
  number = {9},
  pages = {e2020JC016275},
  issn = {2169-9291},
  doi = {10.1029/2020JC016275}
}

@article{Mulia_2021_SyntheticAnalysisEfficacy,
  title = {Synthetic Analysis of the Efficacy of the {{S-net}} System in Tsunami Forecasting},
  author = {Mulia, Iyan E. and Satake, Kenji},
  year = 2021,
  journal = {Earth, Planets and Space},
  volume = {73},
  number = {1},
  pages = {36},
  issn = {1880-5981},
  doi = {10.1186/s40623-021-01368-6}
}

@article{Mungov_2013_DARTTsunameterRetrospective,
  title = {{{DART}}\textregistered{} Tsunameter Retrospective and Real-Time Data: {A} Reflection on 10~Years of Processing in Support of Tsunami Research and Operations},
  author = {Mungov, George and Ebl{\'e}, Marie and Bouchard, Richard},
  year = 2013,
  journal = {Pure and Applied Geophysics},
  volume = {170},
  number = {9},
  pages = {1369--1384},
  issn = {1420-9136},
  doi = {10.1007/s00024-012-0477-5}
}

@article{Nakamura_2012_FDMSimulationSeismicWave,
  title = {{FDM} simulation of seismic-wave propagation for an aftershock of the 2009 {Suruga Bay} earthquake: {Effects} of ocean-bottom topography and seawater layer},
  author = {Nakamura, Takeshi and Takenaka, Hiroshi and Okamoto, Taro and Kaneda, Yoshiyuki},
  year = 2012,
  journal = {Bulletin of the Seismological Society of America},
  volume = {102},
  number = {6},
  pages = {2420--2435},
  issn = {0037-1106},
  doi = {10.1785/0120110356}
}

@article{Noguchi_2016_OceaninfluencedRayleighWaves,
  title = {Ocean-Influenced {{Rayleigh}} Waves from Outer-Rise Earthquakes and Their Effects on Durations of Long-Period Ground Motion},
  author = {Noguchi, Shinako and Maeda, Takuto and Furumura, Takashi},
  year = 2016,
  journal = {Geophysical Journal International},
  volume = {205},
  number = {2},
  pages = {1099--1107},
  issn = {0956-540X},
  doi = {10.1093/gji/ggw074}
}

@article{Oeser_2006_Cluster,
  title = {Cluster design in the {Earth Sciences Tethys}},
  author = {Oeser, Jens and Bunge, Hans Peter and Mohr, Marcus},
  year = 2006,
  journal = {Lecture Notes in Computer Science (including subseries Lecture Notes in Artificial Intelligence and Lecture Notes in Bioinformatics)},
  volume = {4208 LNCS},
  pages = {31--40},
  publisher = {Springer Verlag},
  issn = {16113349},
  doi = {10.1007/11847366_4},
  isbn = {3540393684}
}

@article{Olson_1997_3D_Landers,
  title = {Three-dimensional dynamic simulation of the 1992 {Landers} earthquake},
  author = {Olsen, K. B. and Madariaga, R. and Archuleta, R. J.},
  year = 1997,
  journal = {Science},
  volume = {278},
  number = {5339},
  pages = {834--838},
  publisher = {American Association for the Advancement of Science},
  issn = {00368075},
  doi = {10.1126/SCIENCE.278.5339.834}
}

@article{Pelties_2014_VerificationADERDGMethod,
  title = {Verification of an {{ADER-DG}} Method for Complex Dynamic Rupture Problems},
  author = {Pelties, C and Gabriel, A.-A. and Ampuero, J.-P.},
  year = 2014,
  journal = {Geoscientific Model Development},
  volume = {7},
  number = {3},
  pages = {847--866},
  doi = {10.5194/gmd-7-847-2014}
}

@article{Ragon_2018_AccountingUncertainFault,
  title = {Accounting for Uncertain Fault Geometry in Earthquake Source Inversions -- {{I}}: Theory and Simplified Application},
  author = {Ragon, Th{\'e}a and Sladen, Anthony and Simons, Mark},
  year = 2018,
  journal = {Geophysical Journal International},
  volume = {214},
  number = {2},
  pages = {1174--1190},
  issn = {0956-540X},
  doi = {10.1093/gji/ggy187}
}

@article{Ramos_2021_AssessingMarginWideRupture,
  title = {Assessing margin-wide rupture behaviors along the {Cascadia} megathrust with 3-{D} dynamic rupture simulations},
  author = {Ramos, Marlon D. and Huang, Yihe and Ulrich, Thomas and Li, Duo and Gabriel, Alice Agnes and Thomas, Amanda M.},
  year = 2021,
  journal = {Journal of Geophysical Research: Solid Earth},
  volume = {126},
  number = {7},
  pages = {e2021JB022005},
  publisher = {John Wiley \& Sons, Ltd},
  issn = {2169-9356},
  doi = {10.1029/2021JB022005},
  isbn = {10.1029/2021}
}

@article{Ramos_2022_IntroDR,
  title = {Working with dynamic earthquake rupture models: {A} practical guide},
  author = {Ramos, Marlon D. and Thakur, Prithvi and Huang, Yihe and Harris, Ruth A. and Ryan, Kenny J.},
  year = 2022,
  journal = {Seismological Research Letters},
  volume = {93},
  number = {4},
  pages = {2096--2110},
  publisher = {GeoScienceWorld},
  issn = {0895-0695},
  doi = {10.1785/0220220022}
}

@article{Salaree_2021_RelativeTsunamiHazard,
  title = {Relative Tsunami Hazard From Segments of {{Cascadia Subduction Zone}} For {Mw} 7.5--9.2 Earthquakes},
  author = {Salaree, Amir and Huang, Yihe and Ramos, Marlon D. and Stein, Seth},
  year = 2021,
  journal = {Geophysical Research Letters},
  volume = {48},
  number = {16},
  pages = {e2021GL094174},
  issn = {1944-8007},
  doi = {10.1029/2021GL094174}
}

@article{Sallares_2019_UpperplateRigidityDetermines,
  title = {Upper-Plate Rigidity Determines Depth-Varying Rupture Behaviour of Megathrust Earthquakes},
  author = {Sallar{\`e}s, Valent{\'i} and Ranero, C{\'e}sar R.},
  year = 2019,
  journal = {Nature},
  volume = {576},
  number = {7785},
  pages = {96--101},
  publisher = {Nature Publishing Group},
  issn = {1476-4687},
  doi = {10.1038/s41586-019-1784-0}
}

@article{Satake_2003_FaultSlipSeismic,
  title = {Fault Slip and Seismic Moment of the 1700 {{Cascadia}} Earthquake Inferred from {{Japanese}} Tsunami Descriptions},
  author = {Satake, Kenji and Wang, Kelin and Atwater, Brian F.},
  year = 2003,
  journal = {Journal of Geophysical Research: Solid Earth},
  volume = {108},
  number = {B11},
  issn = {2156-2202},
  doi = {10.1029/2003JB002521}
}

@article{Saunders_2018_AugmentingOnshoreGNSS,
  title = {Augmenting onshore {GNSS} displacements with offshore observations to improve slip characterization for {Cascadia Subduction Zone} earthquakes},
  author = {Saunders, Jessie K. and Haase, Jennifer S.},
  year = 2018,
  journal = {Geophysical Research Letters},
  volume = {45},
  number = {12},
  pages = {6008--6017},
  issn = {1944-8007},
  doi = {10.1029/2018GL078233}
}

@article{Small_2025_CombiningMultisiteTsunami,
  title = {Combining multisite tsunami and deformation modeling to constrain slip distributions for the 1700~{{C}}.{{E}}. {Cascadia} earthquake},
  author = {Small, David T. and Melgar, Diego and La Selle, SeanPaul and Meigs, Andrew},
  year = 2025,
  journal = {Bulletin of the Seismological Society of America},
  volume = {115},
  number = {2},
  pages = {431--451},
  issn = {0037-1106},
  doi = {10.1785/0120240218}
}

@inproceedings{Stanzione_2020_FronteraEvolutionLeadership,
  title = {Frontera: {The} Evolution of Leadership Computing at the {{National Science Foundation}}},
  booktitle = {Practice and {{Experience}} in {{Advanced Research Computing}} 2020: {{Catch}} the {{Wave}}},
  author = {Stanzione, Dan and West, John and Evans, R. Todd and Minyard, Tommy and Ghattas, Omar and Panda, Dhabaleswar K.},
  year = 2020,
  series = {{{PEARC}} '20},
  pages = {106--111},
  publisher = {Association for Computing Machinery},
  doi = {10.1145/3311790.3396656},
  isbn = {978-1-4503-6689-2}
}

@incollection{Tang_2017_RealTimeAssessment16,
  title = {Real-time assessment of the 16 {September} 2015 {Chile} tsunami and implications for near-field forecast},
  booktitle = {The {{Chile-2015}} ({{Illapel}}) {{Earthquake}} and {{Tsunami}}},
  author = {Tang, Liujuan and Titov, Vasily V. and Moore, Christopher and Wei, Yong},
  editor = {Braitenberg, Carla and Rabinovich, Alexander B.},
  year = 2017,
  pages = {267--285},
  publisher = {Springer International Publishing},
  doi = {10.1007/978-3-319-57822-4_19},
  isbn = {978-3-319-57822-4}
}

@article{Tanioka_2018_NearfieldTsunamiInundation,
  title = {Near-Field Tsunami Inundation Forecast Method Assimilating Ocean Bottom Pressure Data: {{A}} Synthetic Test for the 2011 {{Tohoku-Oki}} Tsunami},
  author = {Tanioka, Yuichiro and Gusman, Aditya Riadi},
  year = 2018,
  journal = {Physics of the Earth and Planetary Interiors},
  volume = {283},
  pages = {82--91},
  issn = {0031-9201},
  doi = {10.1016/j.pepi.2018.08.006}
}

@article{tu2026tensor,
    title={Accelerating high-order finite element simulations at extreme Scale with {FP64} tensor cores},
    author={Jiqun Tu and Ian Karlin and John Camier and Veselin Dobrev and Tzanio Kolev and Stefan Henneking and Omar Ghattas},
    year={2026},
    journal={arXiv:2603.09038},
    note={To appear in Proceedings of ISC High Performance 2026},
    doi={10.48550/arXiv.2603.09038}
}

@article{venkat2026oed,
   title = {Sensor placement for tsunami early warning via large-scale {Bayesian} optimal experimental design},
   author = {Sreeram Venkat and Stefan Henneking and Omar Ghattas},
   year={2026},
   journal={arXiv:2604.08812},
   doi={10.48550/arXiv.2604.08812}
}

@inproceedings{Uphoff_2017_ExtremeScaleMultiphysics,
  title = {Extreme Scale Multi-Physics Simulations of the Tsunamigenic 2004 {Sumatra} Megathrust Earthquake},
  booktitle = {Proceedings of the {{International Conference}} for {{High Performance Computing}}, {{Networking}}, {{Storage}} and {{Analysis}}},
  author = {Uphoff, Carsten and Rettenberger, Sebastian and Bader, Michael and Madden, Elizabeth H. and Ulrich, Thomas and Wollherr, Stephanie and Gabriel, Alice-Agnes},
  year = 2017,
  pages = {1--16},
  publisher = {ACM},
  doi = {10.1145/3126908.3126948},
  isbn = {978-1-4503-5114-0}
}

@article{Walton_2021_IntegrativeGeologicalGeophysical,
  title = {Toward an integrative geological and geophysical view of {Cascadia Subduction Zone} earthquakes},
  author = {Walton, Maureen A. L. and Staisch, Lydia M. and Dura, Tina and Pearl, Jessie K. and Sherrod, Brian and Gomberg, Joan and Engelhart, Simon and Tr{\'e}hu, Anne and Watt, Janet and Perkins, Jon and Witter, Robert C. and Bartlow, Noel and Goldfinger, Chris and Kelsey, Harvey and Morey, Ann E. and Sahakian, Valerie J. and Tobin, Harold and Wang, Kelin and Wells, Ray and Wirth, Erin},
  year = 2021,
  journal = {Annual Review of Earth and Planetary Sciences},
  volume = {49},
  number = {Volume 49, 2021},
  pages = {367--398},
  issn = {0084-6597, 1545-4495},
  doi = {10.1146/annurev-earth-071620-065605}
}

@article{Wang_2013_HeterogeneousRuptureGreat,
  title = {Heterogeneous Rupture in the Great {{Cascadia}} Earthquake of 1700 Inferred from Coastal Subsidence Estimates},
  author = {Wang, Pei-Ling and Engelhart, Simon E. and Wang, Kelin and Hawkes, Andrea D. and Horton, Benjamin P. and Nelson, Alan R. and Witter, Robert C.},
  year = 2013,
  journal = {Journal of Geophysical Research: Solid Earth},
  volume = {118},
  number = {5},
  pages = {2460--2473},
  issn = {2169-9356},
  doi = {10.1002/jgrb.50101}
}

@article{Wei_2008_RealtimeExperimentalForecast,
  title = {Real-Time Experimental Forecast of the {{Peruvian}} Tsunami of {{August}} 2007 for {{U}}.{{S}}. Coastlines},
  author = {Wei, Yong and Bernard, Eddie N. and Tang, Liujuan and Weiss, Robert and Titov, Vasily V. and Moore, Christopher and Spillane, Michael and Hopkins, Mike and K{\^a}no{\u g}lu, Utku},
  year = 2008,
  journal = {Geophysical Research Letters},
  volume = {35},
  number = {4},
  issn = {1944-8007},
  doi = {10.1029/2007GL032250}
}

@article{Wendt_2009_TsunamisSplayFault,
  title = {Tsunamis and Splay Fault Dynamics},
  author = {Wendt, James and Oglesby, David D. and Geist, Eric L.},
  year = 2009,
  journal = {Geophysical Research Letters},
  volume = {36},
  number = {15},
  issn = {1944-8007},
  doi = {10.1029/2009GL038295}
}

@article{Williamson_2020_NearFieldTsunamiForecasting,
  title = {Toward near-field tsunami forecasting along the {Cascadia Subduction Zone} using rapid {GNSS} source models},
  author = {Williamson, Amy L. and Melgar, Diego and Crowell, Brendan W. and Arcas, Diego and Melbourne, Timothy I. and Wei, Yong and Kwong, Kevin},
  year = 2020,
  journal = {Journal of Geophysical Research: Solid Earth},
  volume = {125},
  number = {8},
  pages = {e2020JB019636},
  issn = {2169-9356},
  doi = {10.1029/2020JB019636}
}

@article{Wilson_Ma_2021_FullyCoupledDR,
  title = {Wedge plasticity and fully coupled simulations of dynamic rupture and tsunami in the {{Cascadia Subduction Zone}}},
  author = {Wilson, Andrew and Ma, Shuo},
  year = 2021,
  journal = {Journal of Geophysical Research: Solid Earth},
  volume = {126},
  number = {7},
  pages = {e2020JB021627},
  issn = {2169-9356},
  doi = {10.1029/2020JB021627}
}

@misc{Wirp_2025_HellenicArcTsunami,
  title = {Hellenic {Arc} Tsunami Generation from {{Mw8}}+ {{3D}} Margin-Wide Dynamic Rupture Earthquake Scenarios},
  author = {Wirp, Sara Aniko and Kutschera, Fabian and Gabriel, Alice-Agnes and Ulrich, Thomas and Bader, Michael and Lorito, Stefano},
  year = 2025,
  number = {10.31223/X52X6M},
  publisher = {EarthArXiv},
  doi = {10.31223/X52X6M}
}

@article{Wirth_2025_EarthquakeProbabilitiesHazards,
  title = {Earthquake Probabilities and Hazards in the {{U}}.{{S}}. {{Pacific Northwest}}},
  author = {Wirth, Erin A. and Frankel, Arthur D. and Sherrod, Brian and Grant, Alex R. and Dunham, Audrey and Stone, Ian P. and Grossman, Julia},
  year = 2025,
  journal = {U.S. Geological Survey Fact Sheet 2025--3050},
  pages = {6},
  doi = {10.3133/fs20253050}
}
